\documentclass[notitlepage,letterpaper,aps,prd,twocolumn,superscriptaddress,
nofootinbib,longbibliography,showpacs]{revtex4-1}
 \pdfoutput=1
\usepackage{amsmath}
\usepackage{amsfonts}
\usepackage{amssymb}
\usepackage[latin1,utf8]{inputenc}
\usepackage{xcolor,cancel}
\usepackage[normalem]{ulem}
\newcommand{\stkout}[1]{\ifmmode\text{\sout{\ensuremath{#1}}}\else\sout{#1}\fi}
\RequirePackage{flushend}
\usepackage{multirow}

\usepackage{natbib}

\usepackage[colorlinks=true, pdfstartview = FitH, bookmarksopen = true, bookmarksopenlevel=2]{hyperref}
\usepackage{hyperref}
\hypersetup{
	colorlinks=true,
	linkcolor=blue,
	filecolor=magenta,      
	urlcolor=blue,
	pdftitle={DS_jibitesh},
	pdfpagemode=FullScreen,
}

\usepackage[all]{hypcap}

\usepackage{latexsym}
\usepackage{graphicx}
\usepackage{color}
\usepackage{subfigure}

\begin{document}

\title{Background evolution and growth of 
structures in interacting dark energy through dynamical system analysis  }
\author{Wompherdeiki Khyllep}
\email{sjwomkhyllep@gmail.com}
\affiliation{Department of Mathematics, North-Eastern Hill University,
	Shillong, Meghalaya 793022, India}
\affiliation{Department of Mathematics,
	St.\ Anthony's College, Shillong, Meghalaya 793001, India}

\author{Jibitesh Dutta}
\email{jibitesh@nehu.ac.in}
\affiliation{Mathematics Division, Department of Basic Sciences and Social
	Sciences, North-Eastern Hill University,  Shillong, Meghalaya 793022, India}
\affiliation{Inter University Centre for Astronomy and Astrophysics, Pune
	411007, India }

	\author{Spyros Basilakos}
 \email{svasil@academyofathens.gr}
\affiliation{National Observatory of Athens, Lofos Nymfon, 11852 Athens,
Greece}
\affiliation{Academy of Athens, Research Center for Astronomy and
Applied Mathematics, Soranou Efesiou 4, 11527, Athens, Greece}

\author{Emmanuel N. Saridakis}
 \email{msaridak@noa.gr}
\affiliation{National Observatory of Athens, Lofos Nymfon, 11852 Athens,
Greece}
\affiliation{CAS Key Laboratory for Researches in Galaxies and Cosmology,
Department of Astronomy, University of Science and Technology of China, Hefei,
Anhui 230026, P.R. China}
\affiliation{School of Astronomy, School of Physical Sciences,
University of Science and Technology of China, Hefei 230026, P.R. China}

\begin{abstract}
We apply the formalism of dynamical system analysis to investigate the evolution 
of interacting dark energy scenarios at the background and perturbation levels 
in a unified way.  Since the resulting dynamical system contains the extra 
perturbation variable related to the matter overdensity, the critical points of 
the background analysis split, corresponding to   different behavior of matter 
perturbations, and hence to stability properties. From the combined analysis, we 
find critical points that describe the non-accelerating matter-dominated 
epoch with the correct growth of matter structure, and the fact that they are 
saddle provides the natural exit from this phase. Furthermore, we find stable 
attractors at late times corresponding to a dark energy-dominated accelerated 
solution with constant matter perturbations, as required by observations. Thus, 
interacting cosmology can describe the matter and dark energy epochs correctly, 
both at the background and perturbation levels, which reveals the 
capabilities of the interaction. 
\end{abstract}


\maketitle

\section{Introduction}\label{sec:intro}

According to cumulative observations of different origins, the Universe is currently at a phase of accelerating expansion. Although the cosmological constant might be the simplest explanation, the corresponding problem and the possibility of a dynamical nature led to two main directions of modification.   The first is to construct extended theories of 
gravity, which recover general relativity at low energies but which in general 
lead to richer cosmological evolution 
\cite{CANTATA:2021ktz,Capozziello:2011et,Cai:2015emx}. The second avenue is to  
introduce a new sector, collectively known as  dark energy  (DE) \cite{Copeland:2006wr,Cai:2009zp}, with suitable properties that  
can trigger acceleration.   
The dynamical form of DE is usually based on scalar fields, with the simplest choice being the quintessence one. Scalar-field models   usually 
appear in the low-energy limit of various high-energy theories, such as  
the string theory \cite{Damour:1994zq}. However, the inability to explain 
various observational issues have led to a plethora of scalar field 
constructions.

Usually, the DE component is assumed to evolve independently, coupled only to gravity and without interactions with the matter components. 
Nevertheless,  in principle, one cannot neglect   possible interactions 
between the DE and the dark matter (DM) component. Interacting 
DE-DM scenarios are capable of  alleviating the cosmic coincidence problem, leading to  late-time accelerated scaling attractors 
\cite{Chimento:2003iea}.   Additionally, more recently it was shown that  
interacting    models    offer possible solutions to the $H_0$ and $\sigma_8$ 
tensions, and moreover they can
	alleviate the tension between cosmic microwave background and cosmic shear 
measurements
\cite{Pourtsidou:2016ico,An:2017crg,Kumar:2017dnp,  
	 Yang:2018uae,DiValentino:2020vvd,
	DiValentino:2020zio, DiValentino:2021izs}.  
As a result, there have appeared many interacting models which exhibit 
interesting cosmological phenomenology 
\cite{Barrow:2006hia,Amendola:2006dg,He:2008tn,Basilakos:2008ae,
	Gavela:2009cy, Chen:2011cy,Pourtsidou:2013nha, 
	Yang:2014hea, Nunes:2014qoa,Faraoni:2014vra,Salvatelli:2014zta,
	Pan:2012ki, Boehmer:2015kta,Nunes:2016dlj,Boehmer:2015sha, 
	Mukherjee:2016shl,  Cai:2017yww,Yang:2017zjs,Santos:2017bqm,
	Pan:2017ent,  
	Grandon:2018uoe,vonMarttens:2018iav, 
	Bonici:2018qli,Li:2019loh,
	Yang:2019bpr,	
 Mamon:2020spa,BeltranJimenez:2020qdu,Lucca:2021eqy,
Konitopoulos:2021eav}  (for 
reviews 
see \cite{Bolotin:2013jpa,Wang:2016lxa}), and have been confronted with detailed 
observational data, such as the Supernovae Type Ia (SNIa), Baryon
	Acoustic Oscillations (BAO), Cosmic Microwave Background (CMB), Dark Energy 
Survey, galaxy clusters, Hubble function measurements, etc
\cite{Abdalla:2007rd,He:2010im,Solano:2011ie,doi:10.1142/S021827181350082X, 
	Costa:2013sva,Pan:2019jqh,Pan:2019gop,Cheng:2019bkh,Pan:2020zza,Yang:2021oxc}.

In order to examine the viability of a cosmological model, one has first to investigate its evolution at the background level. The next necessary step is the detailed analysis of the cosmological perturbations since these provide information on the stability of the model, they allow for a direct confrontation with growth data, and moreover, they offer a way to distinguish between different scenarios that may lead to the same background evolution 
\cite{Dutta:2017wfd,Khyllep:2019odd,Khyllep:2021pcu,Paliathanasis:2021egx}. 
Concerning interacting scenarios, the effect of DE-DM interaction on the growth 
of structures has been analyzed numerically in \cite{CalderaCabral:2009ja}, and 
it was found that it affects the matter clustering  
\cite{Tsujikawa:2012hv,PhysRevD.90.123007}. Hence, one may observe the imprint 
of the interaction on structure formation, compared to the non-interacting 
scenario.

In general,  cosmological models are determined by complicated equations, and 
the order of complexity increases as we shift from the background to the 
perturbation level. Therefore, it is required to use suitable mathematical 
techniques to extract analytical information and be independent of the initial 
conditions and the specific Universe evolution.  One such powerful mathematical 
tool is the theory of dynamical systems analysis. In particular, 
the phase-space  analysis allows to bypass the complexity of the equations and extract information on the global behavior of the system by examining the   corresponding  critical points since  the 
asymptotic behavior of the model is determined by its form and nature.

The dynamical system approach has been    applied in 
the cosmological context at the background level in numerous works
\cite{wainwrightellis1997,Coley:2003mj,Bahamonde:2017ize,Copeland:1997et,
	Gong:2006sp,
	Setare:2008sf, Matos:2009hf,
	Copeland:2009be,Leyva:2009zz,Leon:2010pu,
 Urena-Lopez:2011gxx, 
	 Leon:2013qh,Fadragas:2013ina,Khyllep:2021yyp,
	Zonunmawia:2018xvf,Dutta:2017kch,Dutta:2016bbs}, including 
interacting 
cosmology  \cite{Billyard:2000bh,Chen:2008ft, Boehmer:2008av}. However, 
at the perturbation 
level it has been applied only partially in very few works  
\cite{1992MNRAS.255..701W,Woszczyna:1988sc,PhysRevD.47.738,Bruni:1993da,
	PhysRevD.48.3562,PhysRevD.62.124007,PhysRevD.90.043501}. Only recently
a dynamical system analysis of the background as well as the 
perturbed
system  was performed systematically for the $\Lambda$CDM paradigm  and 
quintessence scenario with exponential potential 
\cite{Basilakos:2019dof,Alho:2019jho,Landim:2019lvl}.  

Due to the significant effects of  DE-DM interaction on both the background evolution and the growth of structure, it is interesting and necessary to perform detailed dynamical system analysis on interacting cosmology at both the background and perturbation levels. In this way, we can determine the growing mode solution/trajectory determining the structure formation independent of the specific initial conditions. Additionally, we can study how the matter perturbations affect the nature of the background solutions and how perturbations evolve during the cosmological epochs described by each critical point. Finally, we can examine the sensitivity of the structure's growth rate on the strength of the interaction term.

With this motivation, in the present work, we will perform a complete 
dynamical system analysis of various interacting scalar field models,  by 
combining the background and perturbation field equations.   
The manuscript is structured as follows: In Section \ref{sec:bfe}, we present 
the field equations of a  general interacting scalar-field scenario, providing 
the 
equations of the  background evolution, as well as the ones determining 
the linear matter perturbations.
Then in Section 
\ref{sec:DSA}  we perform a detailed phase-space analysis of the full 
system for two interacting models. Finally,  in 
Section \ref{sec:conc} we summarize the obtained results.

\section{Interacting Dark Energy}\label{sec:bfe}

In this section, we briefly review cosmology with dark energy-dark matter interaction, using a scalar field to describe the former. The total action of a minimally coupled scalar field in the framework of general relativity is
\begin{equation}\label{eq:action_pertds}
	S=\int d^4x\sqrt{-g}\left[\frac{R}{2 \kappa^2}+\mathcal{L}_{\phi}+\mathcal{L}_m\right] \,,
\end{equation}
where $\kappa^2$ is the gravitational constant,  $g$ is the determinant of the 
metric $g_{\mu\nu}$, $R$ is the Ricci scalar, 
$\mathcal{L}_m$  and $\mathcal{L}_{\phi}$  are respectively  the matter and scalar-field 
Lagrangian. In particular, $\mathcal{L}_\phi$ is given by 
\begin{equation}
	\mathcal{L}_\phi=-\frac{1}{2} g^{\mu\nu}\partial_\mu \phi \partial_\nu \phi- V(\phi)\,,~~~~~(\mu,\nu=0,1,2,3)
\end{equation}
where $V(\phi)$ is the potential for the scalar field $\phi$, and 
$\mathcal{L}_m$  is considered to correspond to a perfect fluid.

Variation of the action with respect to the metric leads to the field 
equations
\begin{eqnarray}
R_{\mu\nu}-\frac{1}{2}g_{\mu \nu}R=\kappa^2 (T_{\mu\nu}^{(\phi)}+T_{\mu\nu}^{(m)})\,, \label{eq:efe_pertds}
\end{eqnarray}
where the scalar-field  energy-momentum tensor $T_{\mu\nu}^{(\phi)}$  is 
given by
\begin{equation}
T_{\mu\nu}^{(\phi)}=\partial_\mu \phi \partial_\nu \phi- g_{\mu\nu} 
\left[-\frac{1}{2}g^{\alpha\beta}\partial_\alpha \phi \partial_\beta \phi 
+V(\phi)\right]\,,
\end{equation}
and the matter energy-momentum tensor $T_{\mu\nu}^{(m)}$   by
\begin{equation}
T_{\mu\nu}^{(m)}=p_m g_{\mu\nu}+(\rho_m+p_m)u_\mu u_\nu\,,
\end{equation}
with $\rho_m$ and $p_m$    the energy density and pressure of the DM 
component respectively. 

In order to quantitatively describe the interaction  between the DM    
and DE component, the total conservation equation   is split as
\begin{equation}
\nabla_\nu T_m^{\mu\nu}=Q_m^\mu \,,~~~\nabla_\nu T_\phi^{\mu\nu}=Q_\phi^\mu \,, \label{eq:cons_pertds}
\end{equation}
where $Q_\phi^0=-Q_m^0\equiv Q$ is the phenomenological descriptor of the 
interaction,
denoting the rate of energy transfer between the interacting components.  

\subsection{Cosmological equations: Background level}\label{sec:back_eqs}

In order to proceed to cosmological applications,  we consider a homogeneous and isotropic spatially flat  Friedmann-Lema\^itre-Robertson-Walker metric of the form
\begin{equation}\label{eq:FRWmetric}
	ds^2=-dt^2+a^2(t)(dx^2+dy^2+dz^2)\,,
\end{equation}
in  cartesian coordinates. Under the above 
metric, the Einstein's field equations \eqref{eq:efe_pertds} provide the two Friedmann 
equations
\begin{eqnarray}
	3H^2 &=&\kappa^2 \left(\rho_m+\rho_\phi\right)\,, \label{eq:frde_pertds}\\
	2\dot{H}+3H^2 &= &-\kappa^2 (p_m+p_\phi)\,,\label{eq:rce_pertds}
\end{eqnarray}
where   $\rho_\phi=\frac{1}{2} \dot{\phi}^2+ V$, $p_\phi=\frac{1}{2} 
\dot{\phi}^2-V$  are the scalar-field energy density and pressure 
respectively with the upper dot denoting derivative with respect to $t$.
Additionally, under the metric \eqref{eq:FRWmetric}, the conservation  equations 
\eqref{eq:cons_pertds} become
\begin{eqnarray}
	\dot{\rho}_\phi+3H (\rho_\phi+p_\phi)&=&Q\,, \label{eq:cont_phi_pertds}\\
	\dot{\rho}_m+3H(1+w) \rho_m&=&-Q\,, \label{eq:cont_dm_pertds}
\end{eqnarray}
where $w\equiv p_m/\rho_m$ is the  equation of state of DM. Hence, one can see 
that  $Q>0$ corresponds to   energy flow from dark matter to dark energy, 
while $Q<0$ corresponds to  energy transfer in the opposite direction.

We can introduce the  density parameters for the two sectors as
\begin{eqnarray}
	\Omega_\phi 
	&\equiv&\frac{\kappa^2\rho_\phi}{3H^2}=\frac{\kappa^2\dot{\phi}^2}{6H^2}+\frac{
		\kappa^2 V}{3H^2} \,, \label{eq:Omp_pertds}\\
	\Omega_m &\equiv& \frac{\kappa^2\rho_m}{3 H^2}\,, \label{eq:Omm_pertds}
\end{eqnarray}
and thus  the Friedmann equation \eqref{eq:frde_pertds} becomes
\begin{align}\label{eq:Om_reln_pertds}
	\Omega_m+\Omega_\phi=1\,. 
\end{align}
Finally, it proves convenient to define the total, effective, equation-of-state parameter  $w_{\rm eff}$ as
\begin{eqnarray}
	w_{\rm eff} &= &\frac{p_\phi+p_m}{\rho_\phi+\rho_m}=\frac{\frac{1}{2}\dot{\phi}^2-V+w \rho_m}{\frac{1}{2}\dot{\phi}^2+V+\rho_m}\,,
\end{eqnarray}
which is related to the deceleration parameter $q$ as $
w_{\rm eff}=\frac{2q-1}{3}$. As usual, in order to have acceleration  one 
requires the condition $w_{\rm eff}<-\frac{1}{3}$.

\subsection{Cosmological equations: Linear perturbation 
	level}\label{sec:pert_eqs}

We can now examine the behavior of the cosmological system at the linear 
perturbation level. We consider scalar perturbations in the  Newtonian gauge, 
namely
\begin{equation}
	ds^2=-(1+2\Phi)dt^2+a^2(1-2\Phi)(dx^2+dy^2+dz^2)\,,
\end{equation}
and since we are interested in the late-time behavior we have ignored the anisotropic stress. We decompose the matter energy density $\bar{\rho}_m$,  the matter four-velocity 
$\bar{u}_\mu$, the scalar field $\bar{\phi}$ and the energy transfer rate 
$\bar{Q}_A$ into background values and perturbations as
\begin{eqnarray}
	&&\bar{\rho}_m = \rho_m+\delta \rho_m,\nonumber\\
	&&{\bar u}_\mu=  u_\mu +\delta 
	u{_\mu},\nonumber\\
	&&\bar{\phi} = \phi +\delta\phi,\nonumber\\
	&&\bar{Q}= Q+\delta Q\,.
\end{eqnarray}

In this work, we focus on the observationally interesting matter perturbations, and thus we will  not consider the DE ones, since the latter can be assumed to have   a high sound speed and thus does not cluster. Therefore, the evolution equations for the DM energy density perturbation 
$\delta=\delta\rho_m/\rho_m$ and the divergence of velocity
perturbations $\theta$ in the Fourier space are 
\cite{Ma:1995ey,Valiviita:2008iv, Saridakis:2021qxb,Saridakis:2021xqy}:
\begin{equation}
 \!\dot{\delta}+\!\left[3H(c_s^2-w)-\frac{Q}{\rho}\right]\delta+(1+w) 
(\theta-3\dot{\Phi}) =-\frac{\delta Q}{\rho},\label{eq:pert_delt_pertds} 
\end{equation}
\begin{equation}
	\!\dot{\theta} 
+\!\left[H(1\!-\!3w)-\frac{Q}{\rho}+\frac{\dot{w}}{1\!+\!w}\right]\theta -k^2 
\Phi -\frac{c_s^2}{1\!+\!w} k^2 \delta=0,\label{eq:pert_thet_pertds}
\end{equation}
where $c_s^2$ is the sound speed of the fluid,  $\theta=a^{-1}ik^j \delta u_j$
is the divergence of the velocity perturbation, and $k^j$ the wavevector 
component. Finally, in order to  analyze the behavior of DM 
perturbations  one has to combine Eqs. \eqref{eq:pert_delt_pertds} and \eqref{eq:pert_thet_pertds}, with the help of the Poisson's equation \cite{Basilakos:2019dof}. Since structures grow in scales much smaller 
than the Hubble radius $H^{-1}$ (i.e. $k\gg aH$),  the Poisson  
equation becomes
\begin{equation}\label{eq:pois_pertds}
	k^2 \Phi = -\frac{3}{2} H^2 \Omega_m \delta\,.
\end{equation}

\section{Dynamical system analysis}\label{sec:DSA}

In this section, we shall perform a full dynamical system analysis in order to investigate interacting cosmology. Without loss of generality, we  will focus
on two well-studied simple interacting models, namely $Q=\alpha H \rho_m$ and 
$Q=\Gamma \rho_m$. Moreover, concerning the matter sector, as usual, we assume 
it to be dust, i.e. with $w=0$, while  
for   the scalar field potential, we focus on the usual 
exponential potential $V (\phi) = V_0 e^{-\lambda \phi}$, with $V_0>0$ and $\lambda$ is a dimensionless parameter. 
The dynamical system analysis at the background level  has been performed in \cite{Billyard:2000bh,Boehmer:2008av}, 
however, in the present work, we extend the analysis by taking into 
account the effect of perturbations.

\subsection{Interacting model I: $Q=\alpha H \rho_m$}\label{sec:DSA1}

The simple interacting model with $Q=\alpha H  \rho_m$, where $\alpha$ is the 
dimensionless model parameter, has been mathematically designed  to provide 
accelerated scaling attractors, which can alleviate the coincidence problem 
\cite{Billyard:2000bh,Boehmer:2008av}. The sign of  $\alpha$ determines the 
energy transfer direction, i.e. $\alpha>0$ corresponds to  energy flow from   
matter to DE, and vice versa.

Inserting $Q=\alpha H \rho_m$ into  \eqref{eq:pert_delt_pertds} leads to   
$\delta Q = \alpha  \rho_m\delta H + \alpha H \delta \rho_m$, and  $\delta H$ 
is then expressed in terms of $\Phi$. Nevertheless, 
 as it was found in 
\cite{Valiviita:2008iv,He:2008si,2009,Gavela:2009cy,Gavela:2010tm,Yang:2014hea} 
the consideration of  $\delta H$ terms does not have a significant quantitative 
difference in matter overdensity evolution (which is the observable that we are 
interested in in the present work), in  comparison 
to the case where  $\delta H$ terms are neglected. Hence, in the following we 
do not consider these terms. Therefore, the evolution of DM perturbations  obtained from equations 
\eqref{eq:pert_delt_pertds}, \eqref{eq:pert_thet_pertds} and 
\eqref{eq:pois_pertds} can be approximated by
\begin{equation}
	\ddot{\delta} +(2 +3\alpha ) H \dot{\delta} -\frac{3}{2} \Omega_m  H^2 
	\delta=0\,.
\end{equation}

We now proceed to the investigation of perturbations by considering the above 
perturbed equation  along with the background equations 
\eqref{eq:frde_pertds}-\eqref{eq:cont_dm_pertds}. Since we are interested in the qualitative 
behavior of $\delta$, we recast the equations into a first-order autonomous 
system by considering the following auxiliary variables 
\cite{Basilakos:2019dof}:
\begin{eqnarray}
	x=\frac{\kappa \dot{\phi}}{\sqrt{6} H}, ~~~y=\frac{\kappa \sqrt{V}}{\sqrt{3} H}, 
	~~~U=\frac{d (\ln\delta)}{d(\ln a)}
\,. \label{eq:var_pertds}
\end{eqnarray}
Note that  the variables $x, y$ correspond to the background behavior 
of the Universe, while variable $U$ is the usual growth rate which quantities 
the perturbation growth. A 
positive $U$ indicates that inhomogeneities grow, while  negative $U$ indicates 
inhomogeneities decay  whenever perturbation $\delta$ is positive. In terms of 
the above variables, the background cosmological quantities $\Omega_\phi$, 
$\Omega_m$ and $w_{\rm eff}$ can be written as
\begin{eqnarray}
	&&\Omega_\phi=x^2+y^2,\nonumber\\
	&&\Omega_m=1-(x^2+y^2),\nonumber\\
	&&w_{\rm 
eff}=x^2-y^2\,. 
\end{eqnarray}
Hence, under the variables \eqref{eq:var_pertds}, the cosmological equations of the 
present scenario can be expressed in the form of the following dynamical system
\begin{eqnarray}
	x'&=& -3x+\frac{\sqrt{6}}{2} \lambda y^2+\frac{3}{2} x 
(1+x^2-y^2)\nonumber\\
&&+\alpha \frac{(1-x^2-y^2)}{2x}\,, \label{eq:x_1_pertds}\\
	y'&=& -\frac{\sqrt{6}}{2} \lambda xy+\frac{3}{2} y (1+x^2-y^2)\,, \label{eq:y_1_pertds}\\
	U'&=& -U (U+2+3\alpha) +\frac{3}{2} (1-x^2-y^2)\nonumber\\
&& +\frac{3}{2} (1+x^2-y^2) 
U\,, \label{eq:u_1_pertds}
\end{eqnarray}
where primes denote derivatives with respect to $\ln a$ (note that in this 
notation we have simply that $U=\frac{\delta^\prime}{\delta}$).

Since we study the expanding  universe and since the system 
\eqref{eq:x_1_pertds}-\eqref{eq:u_1_pertds} is invariant under a transformation $y \to -y$, 
we focus  only on the phase-space region $y\geq0$. Additionally,  
from the physical condition $0 \leq \Omega_m \leq 1$, the background variables 
$x$ and $y$ are restricted within the circle $x^2+y^2=1$. In summary, the 
background  phase space $\mathbb{B}$  consists of the variables $x, y$, while 
the perturbation phase space $\mathbb{P}$ consists of  $U$, and hence, the  
phase space of the system \eqref{eq:x_1_pertds}-\eqref{eq:u_1_pertds} is the product space 
$\mathbb{B} \times \mathbb{P}$ given by
\begin{eqnarray}
	&&\mathbb{B}\times \mathbb{P}=\left\lbrace (x,y,U) \in \mathbb{R}^2 \times 
	\mathbb{R} :  0 \leq x^2+y^2 \leq 1,\right.\nonumber\\
	&&
\left. \ \ \ \ \ \ \ \ \ \ \ \ \,	
	-1 \leq x \leq 1, 0 \leq y 
\leq 1 
	\right\rbrace \,.
\end{eqnarray}
We mention here that the background equations  \eqref{eq:x_1_pertds},\eqref{eq:y_1_pertds}
on the background space 
$\mathbb{B}$ are decoupled from the   perturbation  equation  \eqref{eq:u_1_pertds} 
on  $\mathbb{P}$, and as usual the projection of an orbit in the product space 
$\mathbb{B} \times \mathbb{P}$ on the background space reduces to the 
corresponding orbit on $\mathbb{B}$.

We proceed to an extraction of critical points of the system 
\eqref{eq:x_1_pertds}-\eqref{eq:u_1_pertds}, by equating the right-hand side of the equations to zero. Then, in order to determine the stability of these points, 
we calculate the eigenvalues of the Jacobian matrix associated with them 
\cite{wainwrightellis1997,Coley:2003mj}. On the physical grounds, a 
stable background point with $U>0$ implies that the matter perturbations grow indefinitely, indicating the instability of the system with respect to matter perturbations. On the contrary, a stable background point with 
$U<0$ implies that the matter perturbations will eventually decay, indicating the asymptotical stability of the system with respect to matter perturbations. 
Finally, when $U=0$ for a stable background point, it implies that the matter 
perturbations of the system asymptotically tend to a fixed value.

	\begin{table*}[!ht]
	\centering
\resizebox{\textwidth}{!}{
		\label{tab:c_pts1_pertds}
	\begin{tabular}{cccccccc}
		\hline\hline
		Point~ &~~ $x$~~ &~~     $y$ ~~& ~~  $U$ 
~~&~~Existence&~~~Stability&~~$\Omega_m$~~ & $w_{\rm 
eff}$      \\
		\hline
			 &      &     & & &Unstable node for $\alpha<1,  
\pm\lambda<\sqrt{6}$&&  \\
$A_{\pm}$ &     $\pm 1$    &   $0$   &$0$&Always&Stable node for $\alpha>3,  
\pm\lambda>\sqrt{6}$&  $0$&   $1$  \\
 &     &     &&&Saddle otherwise& &    \\[2ex]
$B_{\pm}$ &     $\pm 1$     &      $0$& $1-\alpha$& Always &  Unstable node for 
$1<\alpha<3$, $\pm \lambda<\sqrt{6}$ & 0 & 1   
\\
 &     &   & &  &Saddle otherwise&&\\[2ex]
$C$ &     $\frac{\lambda}{\sqrt{6}}$     &      $\sqrt{1-\frac{\lambda^2}{6}}$& 
$0$&$\lambda^2\leq 6$& Stable node for $\alpha>\lambda^2-3$,
$\alpha>\frac{\lambda^2}{2}-2$&$0$  &   $\frac{\lambda^2}{3}-1$  \\ 
&  &&&& Saddle otherwise&&  \\[2ex]

$D$ &     $\frac{\lambda}{\sqrt{6}}$     &      $\sqrt{1-\frac{\lambda^2}{6}}$& 
$\frac{\lambda^2}{2}-\alpha-2$&$\lambda^2 \leq 6$&  Stable node for 
$\alpha>\lambda^2-3$ or
$\alpha<\frac{\lambda^2}{2}-2$&$0$   &   $\frac{\lambda^2}{3}-1$  \\
&  &&&& Saddle otherwise&&  \\[2ex]
$E_{\pm}$ &     $\frac{\alpha+3}{\sqrt{6} \lambda}$     &      
$\frac{\sqrt{(\alpha+3)^2-2 \alpha \lambda^2}}{\sqrt{6} \lambda}$& 
$\ -\frac{1}{4} (\alpha+1)\pm \frac {\sqrt {{\lambda}^{2} \left( 
\alpha+5\right) ^{2}-8\, \left( \alpha+3 \right) ^{2}}}{4\lambda}\ $ &$2 \alpha 
\leq \frac{(\alpha+3)^2}{\lambda^2}\leq \frac{(\alpha+5)^2}{8}$& See Fig. 
\ref{fig:region_E_pertds}  &    
$\frac{(\alpha+3) (\lambda^2-\alpha-3)}{3\lambda^2}$  &   $\frac{\alpha}{3}$  
\\[2ex]
 &  &&&&  $F_+$ stable for $\lambda \sqrt{2 \alpha}>\alpha+3$&     &   
  \\
$F_{\pm}$&     $\frac{\sqrt{\alpha}}{\sqrt{3}}$     &      $0$& $-\frac{1}{4} 
(\alpha+1)\pm \frac{1}{4}\sqrt{(\alpha-3)^2+16}$& $ \alpha \geq 0$ & Saddle 
otherwise& $1-\frac{\alpha}{3}$&  $\frac{\alpha}{3}$\\
&  &&&& $F_-$ saddle always&&  \\[2ex]
&  &&&&  $G_+$ stable for $\lambda \sqrt{2 \alpha}<-\alpha+3$&     &   
\\
$G_{\pm}$&     $-\frac{\sqrt{\alpha}}{\sqrt{3}}$     &      $0$& $-\frac{1}{4} 
(\alpha+1)\pm \frac{1}{4}\sqrt{(\alpha-3)^2+16}$&  $ \alpha \geq 0$  & Saddle 
otherwise& $1-\frac{\alpha}{3}$&  $\frac{\alpha}{3}$\\
&  &&&& $G_-$ saddle always&&  \\
		\hline\hline
	\end{tabular} }
	\caption{The critical points of the system \eqref{eq:x_1_pertds}-\eqref{eq:u_1_pertds}, 
for the interacting model I, namely with $Q=\alpha H \rho_m$, alongside their 
existence and stability conditions, and the 
values of the matter density parameter $\Omega_m$ and the total, effective, 
equation-of-state parameter $w_{\rm 
eff}$.     }
\end{table*}

	\begin{table}[!ht]
	\centering
		\label{tab:eigen1_pertds}
	\begin{tabular}{cccccc}
		\hline\hline
		Point~ &~~ $E_1$~~ &~~     $E_2$ ~~& ~~  $E_3$ ~  \\
		\hline
		$A_{\pm}$ &     $3-\alpha$     &      $3\mp \frac{\sqrt{6}}{2} \lambda$& 
$1-\alpha$    \\
		
		$B_{\pm}$ &     $3-\alpha$     &      $3\mp \frac{\sqrt{6}}{2} \lambda$& 
$\alpha-1$   \\
		
		$C$ &     $\frac{\lambda^2}{2}-3$     &      $\lambda^2-\alpha-3$& 
$\frac{\lambda^2}{2}-\alpha-2$   \\ 
		
		$D$ &    $\frac{\lambda^2}{2}-3$     &      $\lambda^2-\alpha-3$& 
$-\frac{\lambda^2}{2}+\alpha+2$  \\ 
		
		$E_{\pm}$ &     $\Delta_+$     &      $\Delta_-$& $\mp \frac {\sqrt 
{{\lambda}^{2} \left( \alpha+5\right) ^{2}-8\, \left( \alpha+3 \right) 
^{2}}}{2\lambda}$   \\
		
		$F_{\pm}$ &     $\alpha-3$     &      $\frac{1}{2}\left(\alpha+3-\lambda 
\sqrt{2 \alpha}\right)$& $\mp \sqrt{(\alpha-3)^2+16}$   \\
		
		$G_{\pm}$ &     $\alpha-3$     &      $\frac{1}{2}\left(\alpha+3+\lambda 
\sqrt{2 \alpha}\right)$& $\mp \sqrt{(\alpha-3)^2+16}$ 
		\\[1.5ex]
		\hline\hline
	\end{tabular} 
	\caption{The eigenvalues associated with the critical points of the system 
\eqref{eq:x_1_pertds}-\eqref{eq:u_1_pertds}, for the interacting model I, namely with 
$Q=\alpha H \rho_m$. We have defined $\Delta_{\pm}=\frac 
{\lambda(-2\,\alpha\,{\lambda}^{2}+3\,{\alpha}^{2}+6\,
				\alpha-9)\pm \sqrt {S}}{4 \left( \alpha+3 \right) \lambda}$,  
and
$S=4\,{\alpha}^{2}{\lambda}^{6}+4\,{				
\alpha}^{3}{\lambda}^{4}-15\,{\alpha}^{4}{\lambda}^{2}+72\,{\alpha}^{2
			}{\lambda}^{4}+8\,{\alpha}^{5}-204\,{\alpha}^{3}{\lambda}^{2}+180\,
		\alpha\,{\lambda}^{4}+120\,{\alpha}^{4}-882\,{\alpha}^{2}{\lambda}^{2}
	+720\,{\alpha}^{3}-1404\,\alpha\,{\lambda}^{2}+2160\,{\alpha}^{2}-567
\,{\lambda}^{2}+3240\,\alpha+1944$.}
\end{table}

In   Table \ref{tab:c_pts1_pertds}, we summarize the physical critical points of the 
scenario, alongside their existence and stability conditions, as well as the 
values for the observable quantities $\Omega_m$ and   
$w_{\rm eff}$, while in    Table \ref{tab:eigen1_pertds} we give the associated 
eigenvalues. 
 As expected,   the inclusion of scalar perturbations 
	  leads to the split of each  critical point of the background 
analysis into two 
	distinct points, i.e. points that have the same background coordinates $x$ 
and $y$ but different perturbation coordinate $U$.
Hence, the  dynamical system analysis can offer us information on the behavior 
of matter perturbations at these critical points, namely whether they are   
growing,
decaying or remain constant. In particular:
\begin{itemize}
	\item  Points $A_{\pm}$ correspond to DE dominated 
solutions, with a stiff total equation of state, and thus not favored by 
observations, and  with a constant matter perturbation $U=0$. Point $A_+$ is 
unstable node when $\alpha<1$ and $\lambda<\sqrt{6}$, stable when $\alpha>3$ 
and $\lambda> \sqrt{6}$, otherwise it is saddle. Point $A_-$ is unstable node 
when $\alpha<1$ and $\lambda>-\sqrt{6}$, stable when $\alpha>3$ and $\lambda<- 
\sqrt{6}$, and  saddle otherwise.

\item  Points $B_{\pm}$ also correspond to  stiff DE  dominated 
solutions, and thus not favored by 
observations, with the evolution of matter perturbations depending on the 
coupling parameter $\alpha$. For $\alpha<1$, we have a growing mode of 
evolution,   $\alpha>1$ corresponds to  a decaying mode of solution, and 
$\alpha=1$ corresponds to a constant matter perturbation case. Point $B_+$ is 
unstable node when $1<\alpha<3$ and $\lambda<\sqrt{6}$, otherwise it is saddle. 
Similarly, point $B_-$ is unstable node when $1 <\alpha<3$ and 
$\lambda>-\sqrt{6}$, and saddle otherwise.

\item  Point $C$ corresponds to a DE dominated epoch and exists only for $\lambda^2<6$. Its effective equation of state becomes less that 
$-1/3$ for $\lambda^2<2$, giving rise to an accelerating universe.
The point is stable when $\alpha>\lambda^2-3$ and 
$\alpha>\frac{\lambda^2}{2}-2$. Additionally, it has a constant matter 
perturbation i.e. $U=0$. Hence, this point can describe the late-time 
Universe.

\item At the background level, point $D$ coincides with $C$. It corresponds to a DE dominated epoch which is accelerating for  $\lambda^2<2$.
However, when perturbations are considered, point $D$ presents a different 
behavior. In particular,  it can have either growing 
matter perturbations ($U>0$), decaying ($U<0$), or   constant ones ($U=0$ which 
happens   for $\alpha=\frac{\lambda^2}{2}-2$), even when it corresponds to a 
stable late-time DE dominated universe. This makes it the candidate for 
the description of the late-time Universe both at background and perturbation 
levels.

\item Since the eigenvalues corresponding to points  $E_{\pm}$ are complicated, we need to examine their behavior numerically in order to conclude on the stability of $E_{\pm}$. In   Fig. \ref{fig:region_E_pertds} we depict the 
regions in the parameter space that correspond to stable behavior. Note that 
for $\lambda>0$ only point $E_+$ and  for $\lambda<0$ only point 
$E_-$    correspond  to stable  solutions, and within 
these stable regions both points correspond to accelerated solutions if 
$\alpha<-1$.  Concerning the evolution of  matter 
perturbations, in both points $U$ can lie between 0 and 
1, corresponding to growth, for particular parameter regions.  
\begin{figure}[ht]
	\centering
	\includegraphics[width=7cm,height=7cm]{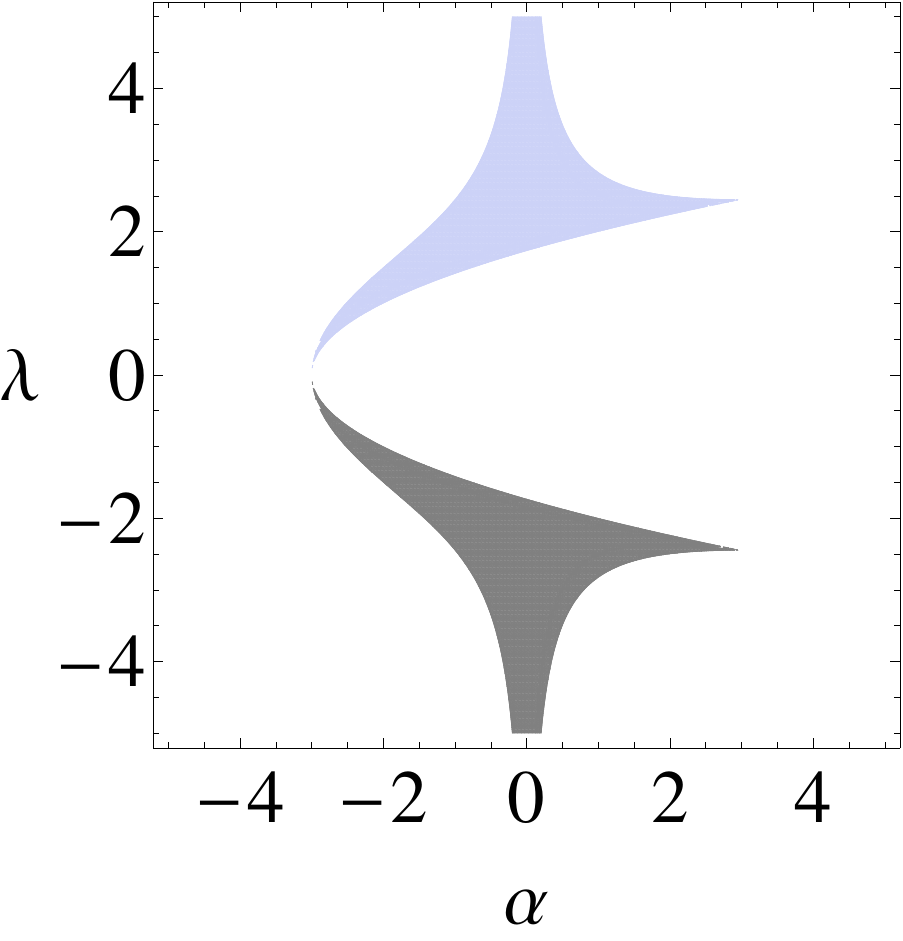}
	\label{fig:region_E_pertds}
	\caption{{\it{Stability regions of points $E_+$ (purple) and $E_-$ (gray) 
				in 
				$(\alpha,\lambda)$ parameter space,  for the interacting model I, namely with 
				$Q=\alpha H \rho_m$.} }}
\end{figure}

\item Points $F_{\pm}$ and $G_{\pm}$ are physical for $0\leq \alpha\leq 3$ 
and they correspond to decelerated scaling solutions. Point $F_+$ is stable when 
$\lambda \sqrt{2 \alpha}>\alpha+3$, otherwise it is saddle.  Point $G_+$ is 
stable when $\lambda\sqrt{2 \alpha}<-\alpha+3$, otherwise it is saddle. It is 
worth 
noting that for small $\alpha$, all four points correspond to matter-dominated solutions at the background level.
Both 
points $F_-$  and $G_-$ are saddle within their physical regions, with decaying matter perturbations. However, points $F_+$ 
and $G_+$ for $\alpha<3$ correspond to growth   of matter 
perturbations, with  $\delta \sim a$. As $\alpha$ increases,  the growth rate is 
smaller than the usual rate during   matter domination, and this reveals the 
effect 
of the coupling towards the structure formation. In summary, taking into 
account both the background and perturbation levels points $F_+$ and 
$G_+$ are the ones that describe the structure formation.

\end{itemize}

		\begin{figure}[ht]
	\centering
	\subfigure[]{%
		
\includegraphics[width=7.6cm,height=7.6cm]{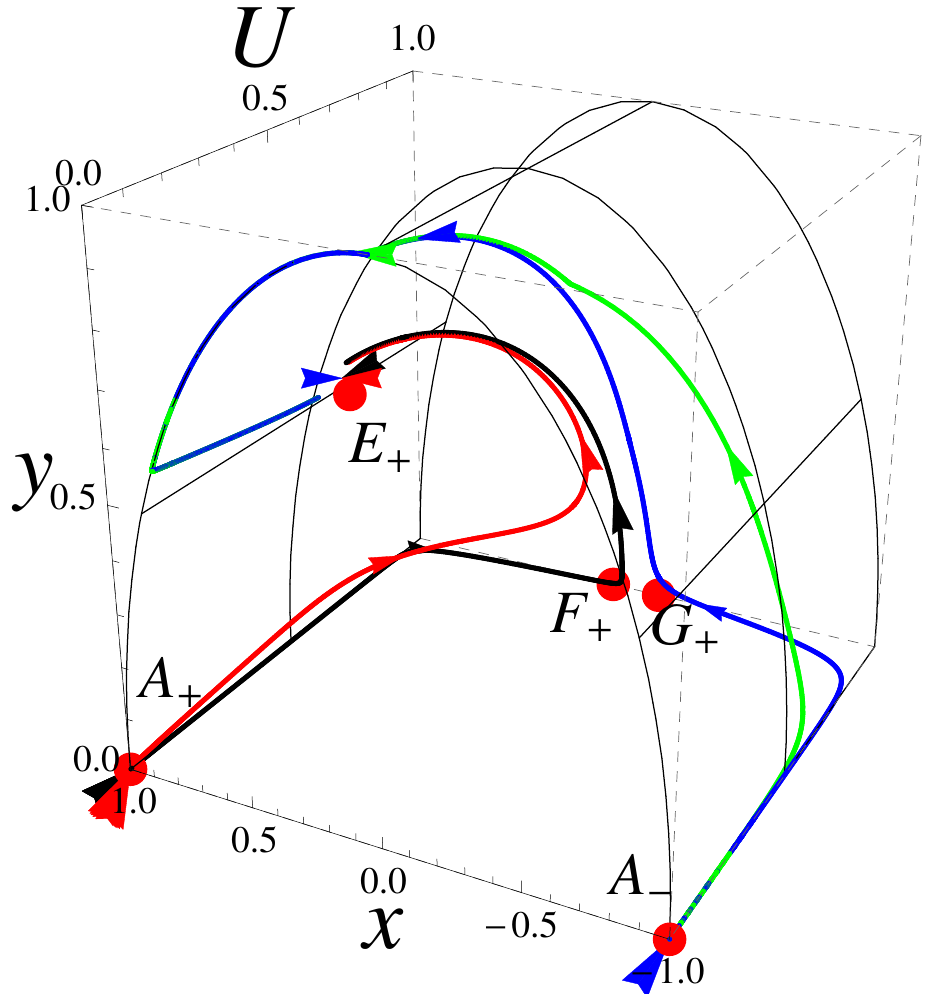}\label{fig:phase1_E_pertds}}
	\qquad 
	\subfigure[]{%
		
\includegraphics[width=7.6cm,height=7.6cm]{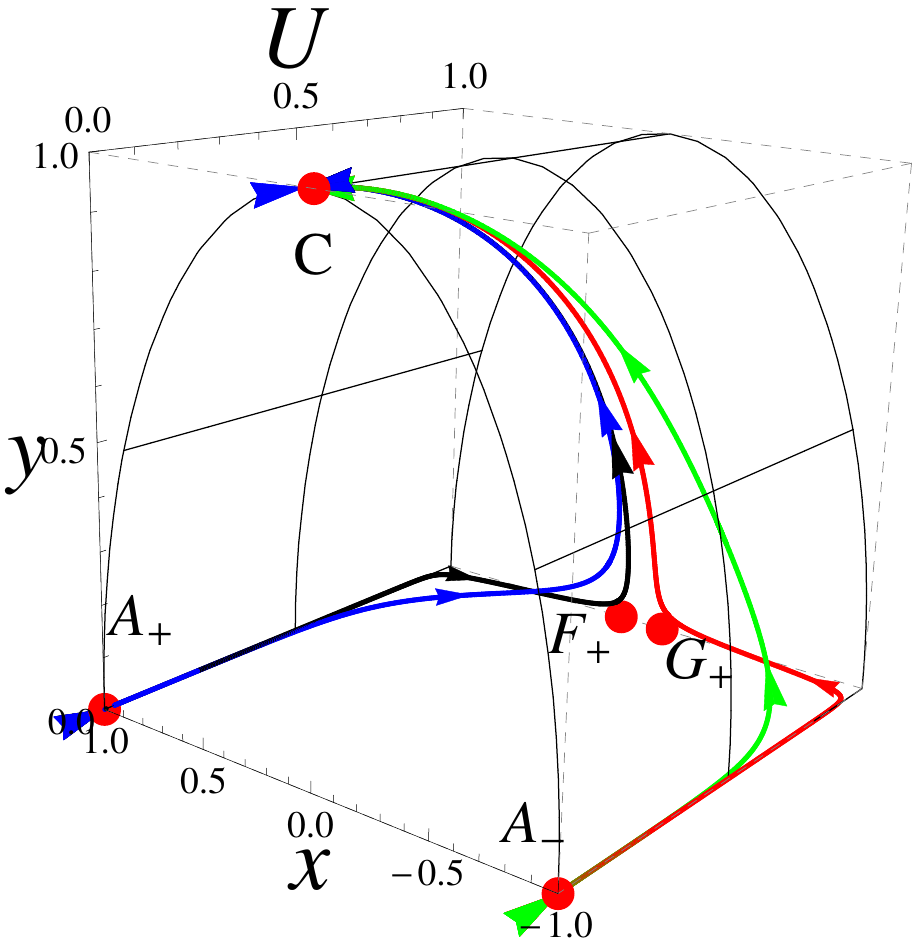}\label{fig:phase1_C_pertds}}
	\caption{{\it{The phase portrait of the system 
\eqref{eq:x_1_pertds}-\eqref{eq:u_1_pertds} of the interacting model I, namely 
with 
$Q=\alpha H \rho_m$. Upper graph:  $\alpha=0.01$, $\lambda=2$, point $E_+$ is 
the attractor.  Lower graph: $\alpha=0.01$, $\lambda=0.1$, point 
$C$ is the attractor.}} }
	\label{fig:phase1_pertds}
\end{figure}

As we observe,  our analysis allows us to view different modes of matter perturbations in terms of critical points.  Points that are the 
same at the background level analysis, they correspond to different behavior of matter perturbations.
From the combined analysis of the 
background and perturbation equations, we find that points $E_-$, $F_+$ and 
$G_+$ are the ones that describe  the non-accelerating 
matter-dominated epoch with the correct growth of matter structure, and the 
fact that they are saddle provides the 
natural exit from this phase. At late times the physically interesting points 
are $C$ and $D$, since they correspond to dark-energy dominated accelerated 
solutions with constant matter perturbations ($C$ always while $D$ for 
$\alpha=\frac{\lambda^2}{2}-2$), as it is required by observations.  Hence, the 
present scenario of interacting cosmology can describe the  thermal history of 
the Universe correctly, both at the background 
and perturbation levels.

In summary, the scenario at hand can   describe both an intermediate epoch with 
the growth of matter perturbations and a late-time accelerating epoch with 
constant matter perturbations, 
offering the correct thermal history of the Universe, at both   background and 
perturbation levels.  This is the main result of the present work and it 
reveals the crucial effect of the interaction.

\begin{figure}[ht]
	\centering
\includegraphics[width=7cm,height=7cm]{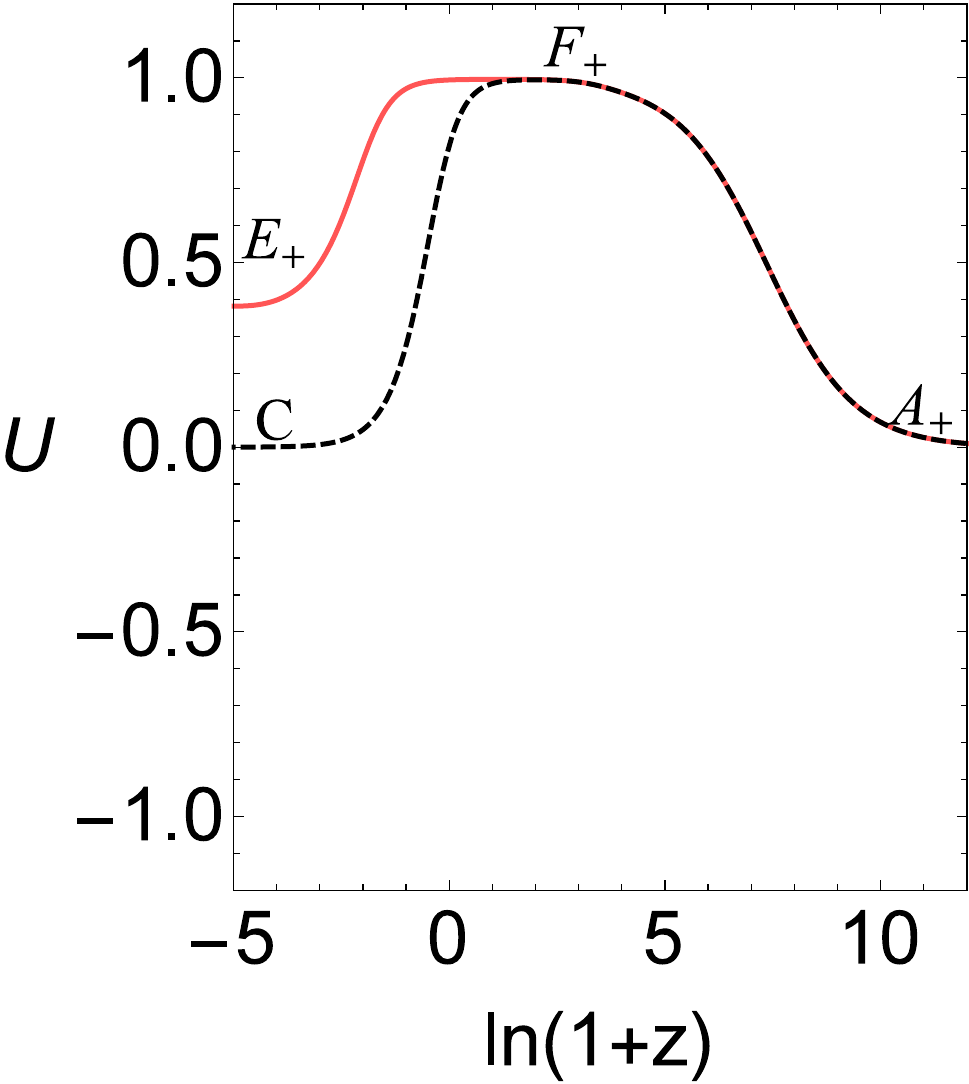}
\label{fig:u_plot_1_pertds}
	\caption{{\it{The evolution of the perturbation quantity $U$ (growth 
rate) for the 
system 
\eqref{eq:x_1_pertds}-\eqref{eq:u_1_pertds} of the interacting model I, namely with 
$Q=\alpha H \rho_m$, with  $\alpha=0.01$ 
and   for $\lambda=2$
(red-solid) and 
$\lambda=0.1$(black-dashed). The red-solid curve corresponds to the transition  
$A_+\to F_+ \to E_+$  and the black-dashed curve to the transition  $A_+\to 
F_+ \to C$.}}}
\end{figure}

In order to be more transparent,  in Fig. \ref{fig:phase1_pertds}  we present 
the phase-space evolution of the system 
\eqref{eq:x_1_pertds}-\eqref{eq:u_1_pertds}, for two 
cases, where we have shown only the growing mode solution of the system. 
 Furthermore,
in Fig. \ref{fig:u_plot_1_pertds}  we 
depict the evolution of the perturbation variable $U$, i.e. the growth rate, for 
two different 
scenarios:  $A_+ \to F_+ \to E_+$ (red-solid curve) describing a 
transition from stiff matter to matter domination and eventually towards a 
decelerated scaling solution, and a sequence $A_+ \to 
F_+ \to C$  (black-dashed curve) describing a transition from a	stiff matter to 
matter domination and eventually towards an  accelerated dark-energy dominated 
solution.  Since $U \geq 0$ at intermediate redshifts,  we 
deduce 
that 
whenever $\delta>0$, $\delta'$ is also non-negative and thus $\delta$ is 
growing 
throughout the evolution, while at late times according to the parameter values 
the growth of perturbation stops and the Universe enters the DE 
dominated epoch. Note that the precise evolution of matter growth depends on 
the interaction parameter $\alpha$.

\begin{figure}[ht]
	\centering
	
\includegraphics[width=7cm,height=5cm]{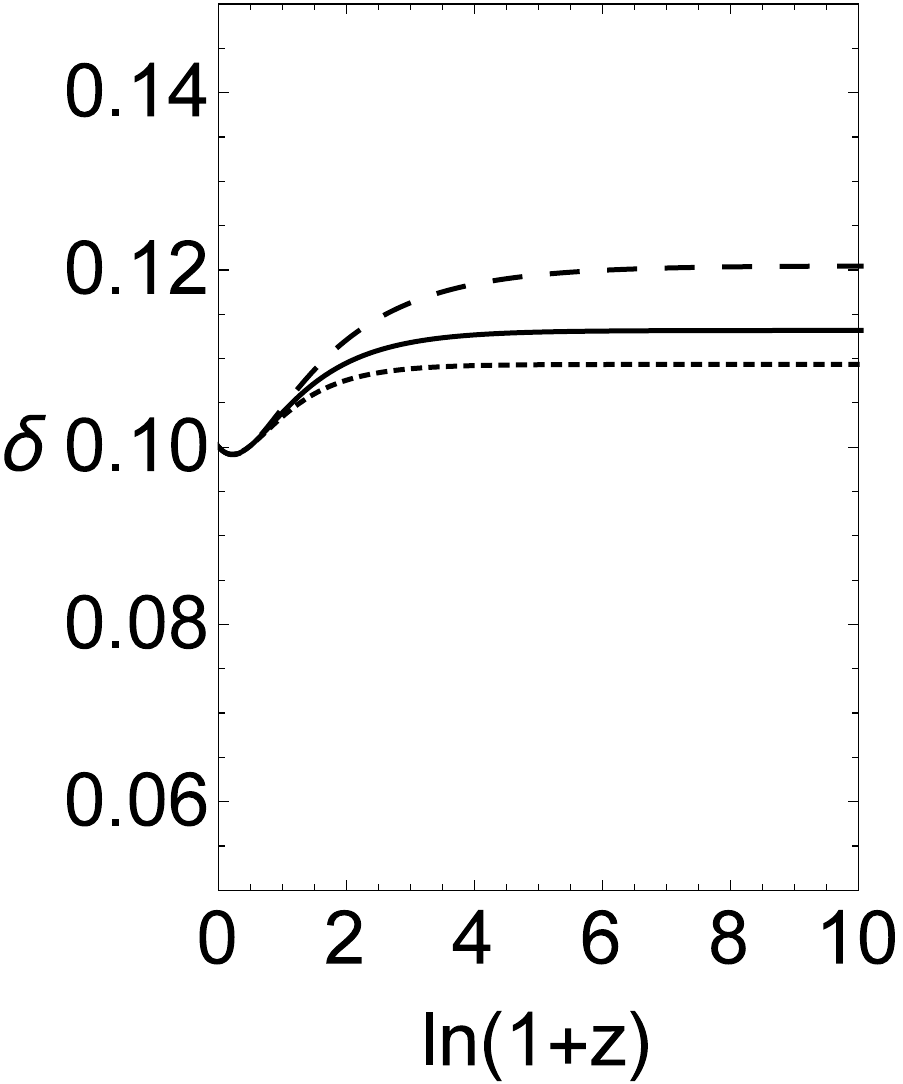}
\label{fig:delta_plot_1_pertds}
	\caption{{\it{The evolution of the  matter overdensity  $\delta$ for the 
interacting model I, namely with $Q=\alpha H \rho_m$, for  $\lambda=0.2$ and 
with $\alpha=-0.1$ (dotted),
$\alpha=0$ (solid), $\alpha=0.1$ (dashed), normalized in the same final value. 
}}}
\end{figure}

Finally, in Fig. 
\ref{fig:delta_plot_1_pertds} we show the evolution of the matter overdensity in the present interacting model and we compare it with the non-interacting case. Imposing the same final conditions in the current universe, we find that for $\alpha<0$ at early times the matter perturbation is smaller compared to the non-interacting case. Therefore,  
the growth rate of structures for   $\alpha<0$   is enhanced in comparison to the non-interacting scenario, which was expected since in this case, DE transforms into DM. On the other hand, the model with $\alpha>0$ exhibits a suppressed structure growth rate.

Lastly, for completeness, we have also examined the possibility of critical 
points at  infinity. Since, all variables apart from $U$ are bounded, we 
consider the transformation $U \to U_\infty$ as $U_\infty=\tan^{-1} U$ with 
$-\frac{\pi}{2} < U_\infty < \frac{\pi}{2}$. Nevertheless, we find that there is 
not any extra critical point at infinity.

\subsection{Interacting model II: $Q=\Gamma \rho_m$}\label{sec:DSA2}

In this subsection, we investigate an interacting model where the energy
transfer rate is determined by the local transfer rate $\Gamma$, which is 
directly associated with the particle or field interactions. This model was used to describe the decay of DM to radiation  
\cite{Cen:2000xv,Oguri:2003nn}, the decay of the curvaton field to radiation 
\cite{Malik:2002jb} and the decay of DM superheavy particles to 
quintessence field  \cite{Ziaeepour:2003qs}. The fact that the interaction rate 
does not depend on $H$, like the model of the previous subsection, as well as 
similar ones, which is a global feature of the Universe, might make the present 
model more physical since the interaction rate is expected to be determined by 
local quantities.
The sign of the constant $\Gamma$ 
determines the energy transfer direction, with the case  $\Gamma>0$ 
corresponding to the decay  of   matter to the scalar field, while the case 
$\Gamma<0$ corresponds to the energy flow in the opposite direction.

\begin{table*}[!ht]
		\centering
\label{tab:c_pts2_pertds}
	\begin{tabular}{ccccccccc}
		\hline\hline
		Point~ &~~ $x$~~ &~~     $y$ ~~& ~~$\zeta$~~&  $U$ 
~~&Existence&Stability&~~$\Omega_m$~~ & 
$w_{\rm eff}$      \\
		\hline
		$\mathcal{A}_{\pm}$ &     $\pm 1$     &  $0$  &  $0$& 
$1$&Always&Saddle&$0$  &   $1$   
\\
		
		$\mathcal{B}_{\pm}$ &   $\pm 1$     &  $0$  &  $0$& 
$0$&Always&Saddle&$0$  &   $1$     
 
		\\
		$\mathcal{C}$ &     $\frac{\lambda}{\sqrt{6}}$    &      
$\sqrt{1-\frac{\lambda^2}{6}}$ &$0$& $0$&$\lambda^2 \leq 6$&Saddle&$0$  &   
$\frac{\lambda^2}{3}-1$   \\ 
		$\mathcal{D}$ &     $\frac{\lambda}{\sqrt{6}}$       &     
$\sqrt{1-\frac{\lambda^2}{6}}$& $0$& $\frac{\lambda^2}{2}-2$&$\lambda^2 \leq 
6$&Saddle&$0$  & $\frac{\lambda^2}{3}-1$  \\
		
		$\mathcal{E}_{\pm}$ &     $\frac{\sqrt{6}}{2 \lambda}$      &  
$\frac{\sqrt{6}}{2 \lambda}$  &    $0$& $-\frac{1}{4} \left(1\pm 
\sqrt{25-\frac{72}{\lambda^2}}\right)$&$\lambda^2 \geq  
\frac{72}{25}$&Saddle&$1-\frac{3}{\lambda^2}$  &   $0$  \\
		
		$\mathcal{F}_{\pm}$ &     $\pm 1$     &      $0$& $1$&$0$ 
&Always&Saddle&   $0$&$1$  
\\
		
			$\mathcal{G}_{\pm}$ &     $\pm 1$     &      $0$& $1$&$1$ 
&Always&Stable node for $\pm \lambda>\sqrt{6}$, $\gamma>0$&     
$0$&$1$\\
	&  &  & & &&Saddle otherwise&   &\\

				$\mathcal{H}$ &     $\frac{\lambda}{\sqrt{6}}$    &       
$\sqrt{1-\frac{\lambda^2}{6}}$ &$1$& $0$&$\lambda^2 \leq 6$&Stable node for $ 
\lambda^2<4$, $\gamma>0$&$0$  &   $\frac{\lambda^2}{3}-1$  \\ 
	&  &  & & &&Saddle otherwise&   &\\
	
			$\mathcal{I}$ &     $\frac{\lambda}{\sqrt{6}}$       &     
$\sqrt{1-\frac{\lambda^2}{6}}$& $1$& $\frac{\lambda^2}{2}-2$&$\lambda^2 \leq 
6$&Stable node for $4< \lambda^2<6$, $\gamma>0$&$0$  &   
$\frac{\lambda^2}{3}-1$ 
 \\ 
&  &  & & &&Saddle otherwise&   &\\
		\hline\hline
	\end{tabular} 
		\caption{The critical points of the system 
\eqref{eq:x_2_pertds}-\eqref{eq:u_2_pertds}, 
for the interacting model II, namely with $Q=\Gamma \rho_m$, alongside their 
existence and stability conditions, and the 
values of the matter density parameter $\Omega_m$ and the total, effective, 
equation-of-state parameter $w_{\rm 
eff}$.    
}
	\end{table*}

\begin{table}[!ht]
	\centering
		\label{tab:eigen2_pertds}
\resizebox{0.5\textwidth}{!}{
	\begin{tabular}{ccccc}
	\hline\hline
	Point~ &~~ $E_1$~~ &~~     $E_2$ ~~& ~~$E_3$~~&  $E_4$     \\
	\hline
	$\mathcal{A}_{\pm}$ &     $3$     &  $3$  &  $3\mp\frac{\sqrt{6} 
\lambda}{2}$& $-1$  \\
	
	$\mathcal{B}_{\pm}$ &       $3$     &  $3$  &  $3\mp\frac{\sqrt{6} 
\lambda}{2}$& $-1$\\
	$\mathcal{C}$ &     $\frac{\lambda^2}{2}$    &      $\lambda^2-3$ 
&$\frac{\lambda^2}{2}-3$& $\frac{\lambda^2}{2}-2$\\
	$\mathcal{D}$ &     $\frac{\lambda^2}{2}$    &      $\lambda^2-3$ 
&$\frac{\lambda^2}{2}-3$& $-\frac{\lambda^2}{2}+2$\\
	
	$\mathcal{E}_{\pm}$ &     $\frac{3}{2}$     &  
$-\frac{3}{4\lambda}\left(\lambda+\sqrt{24-7\lambda^2}\right)$  &   $-\frac{3}{4 
\lambda}\left(\lambda-\sqrt{24-7\lambda^2}\right)$& $\mp 
\frac{\sqrt{25\lambda^2-72}}{2\lambda}$\\
	
	$\mathcal{F}_{\pm}$ &     $-3$     &      $3\mp \frac{\sqrt{6} \lambda}{2}$& 
$-{\rm sgn}(\gamma) \infty$&$1$\\
	
	$\mathcal{G}_{\pm}$ &      $-3$     &      $3\mp \frac{\sqrt{6} 
\lambda}{2}$& $-{\rm sgn}(\gamma) \infty$&$-1$\\
	
	$\mathcal{H}$ &     $-\frac{\lambda^2}{2}$    &      $\frac{\lambda^2}{2}-3$ 
&$-{\rm sgn}(\gamma) \infty$& $\frac{\lambda^2}{2}-2$\\
	$\mathcal{I}$ &      $-\frac{\lambda^2}{2}$    &      
$\frac{\lambda^2}{2}-3$ &$-{\rm sgn}(\gamma) \infty$&  $-\frac{\lambda^2}{2}+2$
	\\[1.5ex]
	\hline\hline
\end{tabular} }
\caption{The eigenvalues associated with the critical points of the system 
\eqref{eq:x_2_pertds}-\eqref{eq:u_2_pertds}, 
for the interacting model II, namely with $Q=\Gamma \rho_m$.  }
\end{table}

Interestingly enough, due to term cancellation the evolution of DM 
perturbations obtained  from Eqs. \eqref{eq:pert_delt_pertds}, \eqref{eq:pert_thet_pertds} 
and \eqref{eq:pois_pertds} for this model is
\begin{equation}\label{eq:delt2_pertds}
	\ddot{\delta} +2 H \dot{\delta} -\frac{3}{2} \Omega_m H^2 \delta=0\,,
\end{equation}
and therefore it coincides  with that of  the non-interacting case. However, 
since the interaction does affect the background evolution, by changing
the evolution of  $H$ and $\Omega_m$ comparing to the  the non-interacting 
case, in the end of the day the 
matter  density   does evolve differently in the interacting and 
non-interacting scenarios. 

In order to transform the cosmological equations into an autonomous form, 
apart from the  variables \eqref{eq:var_pertds} we need to introduce  the additional 
variable $\zeta=\frac{H_0}{H_0+H}$  \cite{Boehmer:2008av}, with  $H_0$ 
the present Hubble constant. Under the variables \eqref{eq:var_pertds} and 
$\zeta$, the 
equations of the present model can be expressed as the following 
dynamical system
\begin{eqnarray}
	x'&=& -3x+\frac{\sqrt{6}}{2} \lambda y^2+\frac{3}{2} x (1+x^2-y^2)
	\nonumber\\
	&&
	-\gamma 
\frac{(1-x^2-y^2) \zeta}{2x(\zeta-1)}\,, \label{eq:x_2_pertds}\\
	y'&=& -\frac{\sqrt{6}}{2} \lambda xy+\frac{3}{2} y (1+x^2-y^2)\,, \label{eq:y_2_pertds}\\
	\zeta'&=& \frac{3}{2} \zeta (1-\zeta) (1+x^2-y^2)\,, \label{eq:z_2_pertds}\\
	U'&=& -U (U+2) +\frac{3}{2} (1-x^2-y^2) 	\nonumber\\
	&&+\frac{3}{2} (1+x^2-y^2) U\,, \label{eq:u_2_pertds}
\end{eqnarray}
where $\gamma=\frac{\Gamma}{H_0}$.
For an expanding Universe ($H>0$) the variable $\zeta$ lies between $0$ and 
$1$. Therefore,   the phase space of the system 
\eqref{eq:x_2_pertds}-\eqref{eq:u_2_pertds} is the product space of the background  phase 
space $\mathbb{B}$ consisting of the variables $x, y, \zeta$, and perturbation 
phase space $\mathbb{P}$ with variable $U$, given by
\begin{eqnarray}
	&&\mathbb{B}\times \mathbb{P}=\left\lbrace (x,y,\zeta,U) \in \mathbb{R}^3 
\times 
	\mathbb{R} :  0 \leq x^2+y^2 \leq 1, \right.\nonumber\\
	&&
\left. \ \ \ \ \ \ \ \ \ \ \ \ \,	-1 \leq x \leq 1, 0 \leq y \leq 1, 0 
\leq \zeta 
	\leq 1 \right\rbrace \,.
\end{eqnarray}

We extract the critical points of the system    
\eqref{eq:x_2_pertds}-\eqref{eq:u_2_pertds}, and we determine their features and stability by 
examining the sign of the corresponding eigenvalues. In Table \ref{tab:c_pts2_pertds}, 
we summarize the physical critical points, alongside the values for the 
observable quantities $\Omega_m$ and   
$w_{\rm eff}$, and  in    Table \ref{tab:eigen2_pertds} we provide the associated 
eigenvalues. In particular: 

\begin{itemize}
	\item  Points $\mathcal{A}_{\pm}$ and $\mathcal{B}_{\pm}$  correspond to 
	stiff DE dominated solutions. At the background level these points 
	are unstable nodes, but including the perturbation level  the points become
	saddle. Moreover, by considering linear 
	perturbations, points $\mathcal{A}_{\pm}$ show a growth of matter perturbations 
	of the form  $\delta \sim a$ even though $\Omega_m=0$.   
	
	\item  Points $\mathcal{C}$ and $\mathcal{D}$ correspond to DE 
	dominated solutions and are always saddle. Their effective equation of state 
	becomes less that 
	$-1/3$ for $\lambda^2<2$, giving rise to an accelerating universe. 
	Nevertheless, although  the two points coincide at the background level, at 
	the level of perturbations they differ, and in particular point $\mathcal{C}$ 
	has constant matter 
	perturbations i.e. $U=0$, while point  $\mathcal{D}$  corresponds to either 
	growth or decay of matter 
	perturbations.
	
	\item  Points $\mathcal{E}_{\pm}$ correspond to scaling matter-dominated, 
	non-accelerating solutions. At the perturbation level, the growth rate 
	of  matter perturbations are smaller than the standard matter-dominated 
	epoch. Point $\mathcal{E}_{+}$ corresponds to decaying matter perturbations. 
	However, point  $\mathcal{E}_-$ for $\lambda>3$ corresponds to the growth of matter perturbations, and especially for large values of $\lambda$ it describes a matter-dominated era with matter overdensity evolving as $\delta \sim a$.
	
	\item Points $\mathcal{F}_{\pm}$ and $\mathcal{G}_{\pm}$  correspond to 
	stiff DE dominated solutions. Points $\mathcal{F}_{\pm}$ are always saddle but points $\mathcal{G}_{\pm}$ can be either saddle or stable node depending on $\lambda$ and the sign of $\gamma$. At the perturbation level, 
	while points $\mathcal{F}_{\pm}$  correspond to a constant growth rate of 
matter perturbations, points $\mathcal{G}_{\pm}$  correspond to unstable growth. 
Finally, note that while in the pure background analysis points 
$\mathcal{F}_{\pm}$ can be stable,   the inclusion of perturbations make them 
saddle.

	\item  Points $\mathcal{H}$ and $\mathcal{I}$ correspond to DE 
	dominated solutions, which for  $\lambda^2<2$ exhibit acceleration.
	Point $\mathcal{H}$ is stable at both the background and 
	perturbation level when $\gamma>0$ and  $\lambda^2<4$, and similarly point 
	$\mathcal{I}$ is stable when $\gamma>0$ and $\lambda^2>4$. 
	Point $\mathcal{H}$ has a constant growth rate of matter 
	perturbations, while point $\mathcal{I}$ corresponds to   decay   of 
matter 
	perturbations in its stability region, while it corresponds to unstable 
	perturbation growth when it is a saddle.

\end{itemize}

		\begin{figure}[ht]
		\centering		
\includegraphics[width=7cm,height=7cm]{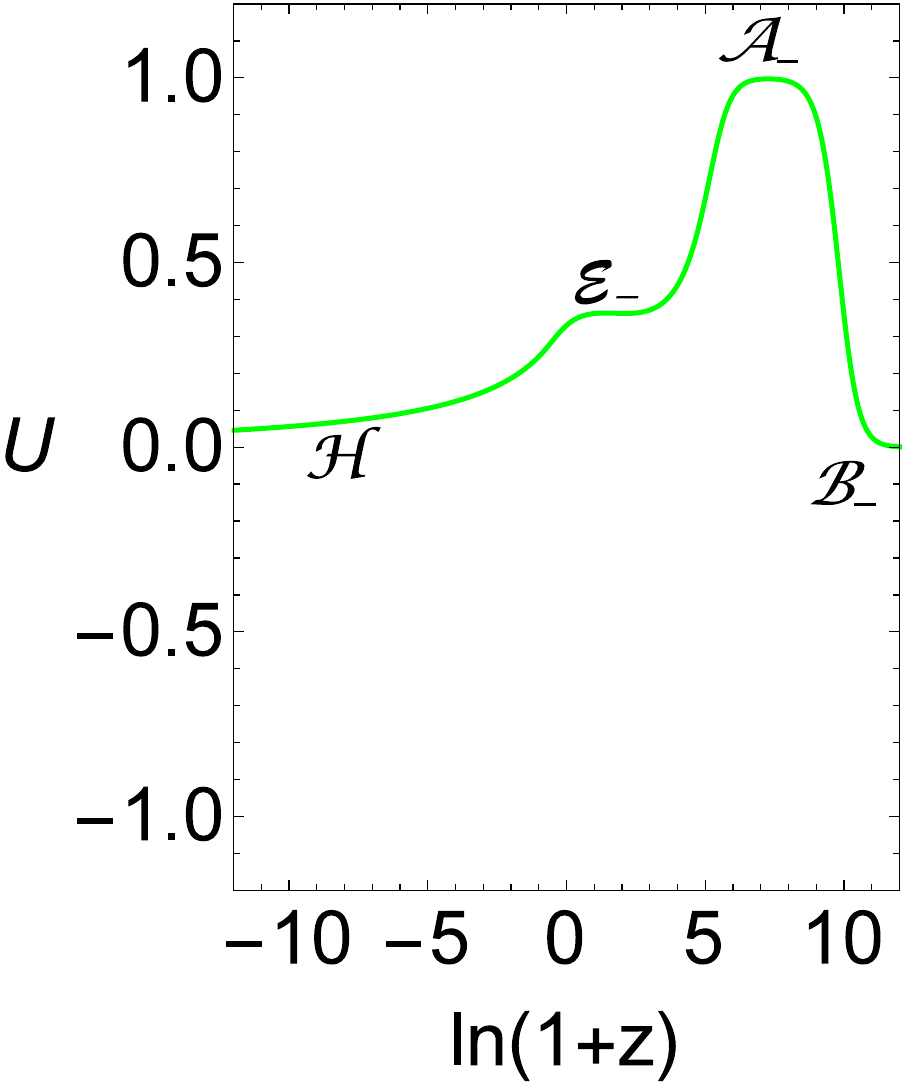}\label{fig:u_plot_2_pertds}
		\caption{{\it{The evolution of the perturbation quantity $U$ (growth 
rate) for the 
system 	\eqref{eq:x_2_pertds}-\eqref{eq:u_2_pertds} of the 
interacting model II,  namely with 
					$Q=\Gamma \rho_m$,   with    $\lambda^2=3.9$, and for     
					$\gamma=10^{-5}$.  It corresponds to the 
transition  $\mathcal{B}_-\to \mathcal{A}_- \to \mathcal{E}_-\to \mathcal{H}$.}}}
	\end{figure}

Similarly to the previous interacting model,  we see that the inclusion of 
perturbations allow us to   differentiate amongst critical points that are 
equivalent at the background level. Additionally, performing an analysis at 
infinity, we find 
that there are no extra critical points. From the combined analysis of the 
background and perturbation level, we find that point $\mathcal{E}_{-}$ is 
 physically interesting for the description of the epoch where matter 
perturbations are generated, and the fact that it is saddle gives the 
natural exit from this phase. At late times the physically interesting point is 
$\mathcal{H}$, which corresponds to  DE dominated accelerated 
solution with constant matter perturbations, in agreement with observations. 
 For completeness, in Fig. 
\ref{fig:u_plot_2_pertds} we depict the evolution  of the growth rate $U$,
which shows that for a wide range of initial conditions, $\delta$ increases 
at intermediate redshifts and then it asymptotically approaches a constant 
value (since $U$ goes to zero).  Moreover,  the precise evolution of matter 
growth depends on 
the interaction parameter $\gamma$ as expected. 

\section{Conclusions}\label{sec:conc}

Cosmological scenarios with interactions between DM and DE 
sectors are widely studied since they offer alleviation of the cosmic 
coincidence problem, and additionally, they may lead to solutions to the Hubble 
and $\sigma_8$ tensions. Hence, in the present work, we applied the formalism 
of dynamical system analysis to investigate the evolution of 
interacting scenarios both at the background and perturbation levels in a 
unified way.

We transformed the background and perturbation equations into an autonomous system and investigated them for two interaction forms. Since the resulting    dynamical system contains the  extra perturbation
variable  related to the matter overdensity, the critical points of the 
background   analysis split into more points, corresponding to different behavior 
of matter perturbations, and hence to stability properties.

For both models, we obtained critical points describing a wide range of 
interesting cosmological solutions, corresponding to DM, scaling, or DE dominated ones, accelerating and non-accelerating, with decaying, constant, or growing matter perturbations.
In particular, from the combined analysis of the background and perturbation 
equations, we found points that describe the non-accelerating matter-dominated 
epoch with the correct growth of matter structure. The fact that they are saddle 
provides the natural exit from this phase. Furthermore, at late times, we 
found stable attractors that correspond to a DE-dominated accelerated solution 
with constant matter perturbations, as observations require it.

In summary, interacting cosmology, and more efficiently interacting 
model I, can correctly describe the matter and DE epochs at the background and 
perturbation levels. Moreover,    we have an extra parameter to adjust 
both the background and perturbation 
behavior, namely the coupling parameter. Hence, for suitable parameter regions, 
one can obtain 
the correct thermal history of the Universe, at both   background 
and perturbation levels,   offering a matter-dominated phase  where
matter perturbations grow and which ends naturally, and then the 
transition to the DE-dominated accelerating phase. These features reveal the 
significant effect of the interaction.

It would be interesting to investigate the phase space of both the 
background and perturbation levels in interacting scenarios in which dark 
energy has an effective origin, namely instead of quintessence field to arise 
from modified gravity. Additionally, one could also   apply the 
dynamical system analysis in the case 
of tensor perturbations. These studies lie beyond the scope of the present work 
and are left for future projects.

\begin{acknowledgments}
	
	 JD was supported by the Core Research Grant of SERB, Department of Science 
and Technology India (File No. CRG/2018/001035) and the Associate program of 
IUCAA.   SB and ENS would like to acknowledge the contribution of the COST 
Action 
CA18108 ``Quantum Gravity Phenomenology in the multi-messenger approach''.
\end{acknowledgments}

\bibliography{pert_ds}

\begin{thebibliography}{103}%
\makeatletter
\providecommand \@ifxundefined [1]{%
 \@ifx{#1\undefined}
}%
\providecommand \@ifnum [1]{%
 \ifnum #1\expandafter \@firstoftwo
 \else \expandafter \@secondoftwo
 \fi
}%
\providecommand \@ifx [1]{%
 \ifx #1\expandafter \@firstoftwo
 \else \expandafter \@secondoftwo
 \fi
}%
\providecommand \natexlab [1]{#1}%
\providecommand \enquote  [1]{``#1''}%
\providecommand \bibnamefont  [1]{#1}%
\providecommand \bibfnamefont [1]{#1}%
\providecommand \citenamefont [1]{#1}%
\providecommand \href@noop [0]{\@secondoftwo}%
\providecommand \href [0]{\begingroup \@sanitize@url \@href}%
\providecommand \@href[1]{\@@startlink{#1}\@@href}%
\providecommand \@@href[1]{\endgroup#1\@@endlink}%
\providecommand \@sanitize@url [0]{\catcode `\\12\catcode `\$12\catcode
  `\&12\catcode `\#12\catcode `\^12\catcode `\_12\catcode `\%12\relax}%
\providecommand \@@startlink[1]{}%
\providecommand \@@endlink[0]{}%
\providecommand \url  [0]{\begingroup\@sanitize@url \@url }%
\providecommand \@url [1]{\endgroup\@href {#1}{\urlprefix }}%
\providecommand \urlprefix  [0]{URL }%
\providecommand \Eprint [0]{\href }%
\providecommand \doibase [0]{http://dx.doi.org/}%
\providecommand \selectlanguage [0]{\@gobble}%
\providecommand \bibinfo  [0]{\@secondoftwo}%
\providecommand \bibfield  [0]{\@secondoftwo}%
\providecommand \translation [1]{[#1]}%
\providecommand \BibitemOpen [0]{}%
\providecommand \bibitemStop [0]{}%
\providecommand \bibitemNoStop [0]{.\EOS\space}%
\providecommand \EOS [0]{\spacefactor3000\relax}%
\providecommand \BibitemShut  [1]{\csname bibitem#1\endcsname}%
\let\auto@bib@innerbib\@empty
\bibitem [{\citenamefont {Saridakis}\ \emph
  {et~al.}(2021{\natexlab{a}})\citenamefont {Saridakis} \emph
  {et~al.}}]{CANTATA:2021ktz}%
  \BibitemOpen
  \bibfield  {author} {\bibinfo {author} {\bibfnamefont {Emmanuel~N.}\
  \bibnamefont {Saridakis}} \emph {et~al.} (\bibinfo {collaboration}
  {CANTATA}),\ }\bibfield  {title} {\enquote {\bibinfo {title} {{Modified
  Gravity and Cosmology: An Update by the CANTATA Network}},}\ }\href@noop {}
  {\  (\bibinfo {year} {2021}{\natexlab{a}})},\ \Eprint
  {http://arxiv.org/abs/2105.12582} {arXiv:2105.12582 [gr-qc]} \BibitemShut
  {NoStop}%
\bibitem [{\citenamefont {Capozziello}\ and\ \citenamefont
  {De~Laurentis}(2011)}]{Capozziello:2011et}%
  \BibitemOpen
  \bibfield  {author} {\bibinfo {author} {\bibfnamefont {Salvatore}\
  \bibnamefont {Capozziello}}\ and\ \bibinfo {author} {\bibfnamefont
  {Mariafelicia}\ \bibnamefont {De~Laurentis}},\ }\bibfield  {title} {\enquote
  {\bibinfo {title} {{Extended Theories of Gravity}},}\ }\href {\doibase
  10.1016/j.physrep.2011.09.003} {\bibfield  {journal} {\bibinfo  {journal}
  {Phys. Rept.}\ }\textbf {\bibinfo {volume} {509}},\ \bibinfo {pages}
  {167--321} (\bibinfo {year} {2011})},\ \Eprint
  {http://arxiv.org/abs/1108.6266} {arXiv:1108.6266 [gr-qc]} \BibitemShut
  {NoStop}%
\bibitem [{\citenamefont {Cai}\ \emph {et~al.}(2016)\citenamefont {Cai},
  \citenamefont {Capozziello}, \citenamefont {De~Laurentis},\ and\
  \citenamefont {Saridakis}}]{Cai:2015emx}%
  \BibitemOpen
  \bibfield  {author} {\bibinfo {author} {\bibfnamefont {Yi-Fu}\ \bibnamefont
  {Cai}}, \bibinfo {author} {\bibfnamefont {Salvatore}\ \bibnamefont
  {Capozziello}}, \bibinfo {author} {\bibfnamefont {Mariafelicia}\ \bibnamefont
  {De~Laurentis}}, \ and\ \bibinfo {author} {\bibfnamefont {Emmanuel~N.}\
  \bibnamefont {Saridakis}},\ }\bibfield  {title} {\enquote {\bibinfo {title}
  {{f(T) teleparallel gravity and cosmology}},}\ }\href {\doibase
  10.1088/0034-4885/79/10/106901} {\bibfield  {journal} {\bibinfo  {journal}
  {Rept. Prog. Phys.}\ }\textbf {\bibinfo {volume} {79}},\ \bibinfo {pages}
  {106901} (\bibinfo {year} {2016})},\ \Eprint
  {http://arxiv.org/abs/1511.07586} {arXiv:1511.07586 [gr-qc]} \BibitemShut
  {NoStop}%
\bibitem [{\citenamefont {Copeland}\ \emph {et~al.}(2006)\citenamefont
  {Copeland}, \citenamefont {Sami},\ and\ \citenamefont
  {Tsujikawa}}]{Copeland:2006wr}%
  \BibitemOpen
  \bibfield  {author} {\bibinfo {author} {\bibfnamefont {Edmund~J.}\
  \bibnamefont {Copeland}}, \bibinfo {author} {\bibfnamefont {M.}~\bibnamefont
  {Sami}}, \ and\ \bibinfo {author} {\bibfnamefont {Shinji}\ \bibnamefont
  {Tsujikawa}},\ }\bibfield  {title} {\enquote {\bibinfo {title} {{Dynamics of
  dark energy}},}\ }\href {\doibase 10.1142/S021827180600942X} {\bibfield
  {journal} {\bibinfo  {journal} {Int. J. Mod. Phys. D}\ }\textbf {\bibinfo
  {volume} {15}},\ \bibinfo {pages} {1753--1936} (\bibinfo {year} {2006})},\
  \Eprint {http://arxiv.org/abs/hep-th/0603057} {arXiv:hep-th/0603057}
  \BibitemShut {NoStop}%
\bibitem [{\citenamefont {Cai}\ \emph {et~al.}(2010)\citenamefont {Cai},
  \citenamefont {Saridakis}, \citenamefont {Setare},\ and\ \citenamefont
  {Xia}}]{Cai:2009zp}%
  \BibitemOpen
  \bibfield  {author} {\bibinfo {author} {\bibfnamefont {Yi-Fu}\ \bibnamefont
  {Cai}}, \bibinfo {author} {\bibfnamefont {Emmanuel~N.}\ \bibnamefont
  {Saridakis}}, \bibinfo {author} {\bibfnamefont {Mohammad~R.}\ \bibnamefont
  {Setare}}, \ and\ \bibinfo {author} {\bibfnamefont {Jun-Qing}\ \bibnamefont
  {Xia}},\ }\bibfield  {title} {\enquote {\bibinfo {title} {{Quintom Cosmology:
  Theoretical implications and observations}},}\ }\href {\doibase
  10.1016/j.physrep.2010.04.001} {\bibfield  {journal} {\bibinfo  {journal}
  {Phys. Rept.}\ }\textbf {\bibinfo {volume} {493}},\ \bibinfo {pages} {1--60}
  (\bibinfo {year} {2010})},\ \Eprint {http://arxiv.org/abs/0909.2776}
  {arXiv:0909.2776 [hep-th]} \BibitemShut {NoStop}%
\bibitem [{\citenamefont {Damour}\ and\ \citenamefont
  {Polyakov}(1994)}]{Damour:1994zq}%
  \BibitemOpen
  \bibfield  {author} {\bibinfo {author} {\bibfnamefont {T.}~\bibnamefont
  {Damour}}\ and\ \bibinfo {author} {\bibfnamefont {Alexander~M.}\ \bibnamefont
  {Polyakov}},\ }\bibfield  {title} {\enquote {\bibinfo {title} {{The String
  dilaton and a least coupling principle}},}\ }\href {\doibase
  10.1016/0550-3213(94)90143-0} {\bibfield  {journal} {\bibinfo  {journal}
  {Nucl. Phys. B}\ }\textbf {\bibinfo {volume} {423}},\ \bibinfo {pages}
  {532--558} (\bibinfo {year} {1994})},\ \Eprint
  {http://arxiv.org/abs/hep-th/9401069} {arXiv:hep-th/9401069} \BibitemShut
  {NoStop}%
\bibitem [{\citenamefont {Chimento}\ \emph {et~al.}(2003)\citenamefont
  {Chimento}, \citenamefont {Jakubi}, \citenamefont {Pavon},\ and\
  \citenamefont {Zimdahl}}]{Chimento:2003iea}%
  \BibitemOpen
  \bibfield  {author} {\bibinfo {author} {\bibfnamefont {Luis~P.}\ \bibnamefont
  {Chimento}}, \bibinfo {author} {\bibfnamefont {Alejandro~S.}\ \bibnamefont
  {Jakubi}}, \bibinfo {author} {\bibfnamefont {Diego}\ \bibnamefont {Pavon}}, \
  and\ \bibinfo {author} {\bibfnamefont {Winfried}\ \bibnamefont {Zimdahl}},\
  }\bibfield  {title} {\enquote {\bibinfo {title} {{Interacting quintessence
  solution to the coincidence problem}},}\ }\href {\doibase
  10.1103/PhysRevD.67.083513} {\bibfield  {journal} {\bibinfo  {journal} {Phys.
  Rev. D}\ }\textbf {\bibinfo {volume} {67}},\ \bibinfo {pages} {083513}
  (\bibinfo {year} {2003})},\ \Eprint {http://arxiv.org/abs/astro-ph/0303145}
  {arXiv:astro-ph/0303145} \BibitemShut {NoStop}%
\bibitem [{\citenamefont {Pourtsidou}\ and\ \citenamefont
  {Tram}(2016)}]{Pourtsidou:2016ico}%
  \BibitemOpen
  \bibfield  {author} {\bibinfo {author} {\bibfnamefont {Alkistis}\
  \bibnamefont {Pourtsidou}}\ and\ \bibinfo {author} {\bibfnamefont {Thomas}\
  \bibnamefont {Tram}},\ }\bibfield  {title} {\enquote {\bibinfo {title}
  {{Reconciling CMB and structure growth measurements with dark energy
  interactions}},}\ }\href {\doibase 10.1103/PhysRevD.94.043518} {\bibfield
  {journal} {\bibinfo  {journal} {Phys. Rev. D}\ }\textbf {\bibinfo {volume}
  {94}},\ \bibinfo {pages} {043518} (\bibinfo {year} {2016})},\ \Eprint
  {http://arxiv.org/abs/1604.04222} {arXiv:1604.04222 [astro-ph.CO]}
  \BibitemShut {NoStop}%
\bibitem [{\citenamefont {An}\ \emph {et~al.}(2018)\citenamefont {An},
  \citenamefont {Feng},\ and\ \citenamefont {Wang}}]{An:2017crg}%
  \BibitemOpen
  \bibfield  {author} {\bibinfo {author} {\bibfnamefont {Rui}\ \bibnamefont
  {An}}, \bibinfo {author} {\bibfnamefont {Chang}\ \bibnamefont {Feng}}, \ and\
  \bibinfo {author} {\bibfnamefont {Bin}\ \bibnamefont {Wang}},\ }\bibfield
  {title} {\enquote {\bibinfo {title} {{Relieving the Tension between Weak
  Lensing and Cosmic Microwave Background with Interacting Dark Matter and Dark
  Energy Models}},}\ }\href {\doibase 10.1088/1475-7516/2018/02/038} {\bibfield
   {journal} {\bibinfo  {journal} {JCAP}\ }\textbf {\bibinfo {volume} {02}},\
  \bibinfo {pages} {038} (\bibinfo {year} {2018})},\ \Eprint
  {http://arxiv.org/abs/1711.06799} {arXiv:1711.06799 [astro-ph.CO]}
  \BibitemShut {NoStop}%
\bibitem [{\citenamefont {Kumar}\ and\ \citenamefont
  {Nunes}(2017)}]{Kumar:2017dnp}%
  \BibitemOpen
  \bibfield  {author} {\bibinfo {author} {\bibfnamefont {Suresh}\ \bibnamefont
  {Kumar}}\ and\ \bibinfo {author} {\bibfnamefont {Rafael~C.}\ \bibnamefont
  {Nunes}},\ }\bibfield  {title} {\enquote {\bibinfo {title} {{Echo of
  interactions in the dark sector}},}\ }\href {\doibase
  10.1103/PhysRevD.96.103511} {\bibfield  {journal} {\bibinfo  {journal} {Phys.
  Rev. D}\ }\textbf {\bibinfo {volume} {96}},\ \bibinfo {pages} {103511}
  (\bibinfo {year} {2017})},\ \Eprint {http://arxiv.org/abs/1702.02143}
  {arXiv:1702.02143 [astro-ph.CO]} \BibitemShut {NoStop}%
\bibitem [{\citenamefont {Yang}\ \emph
  {et~al.}(2018{\natexlab{a}})\citenamefont {Yang}, \citenamefont {Mukherjee},
  \citenamefont {Di~Valentino},\ and\ \citenamefont {Pan}}]{Yang:2018uae}%
  \BibitemOpen
  \bibfield  {author} {\bibinfo {author} {\bibfnamefont {Weiqiang}\
  \bibnamefont {Yang}}, \bibinfo {author} {\bibfnamefont {Ankan}\ \bibnamefont
  {Mukherjee}}, \bibinfo {author} {\bibfnamefont {Eleonora}\ \bibnamefont
  {Di~Valentino}}, \ and\ \bibinfo {author} {\bibfnamefont {Supriya}\
  \bibnamefont {Pan}},\ }\bibfield  {title} {\enquote {\bibinfo {title}
  {{Interacting dark energy with time varying equation of state and the $H_0$
  tension}},}\ }\href {\doibase 10.1103/PhysRevD.98.123527} {\bibfield
  {journal} {\bibinfo  {journal} {Phys. Rev. D}\ }\textbf {\bibinfo {volume}
  {98}},\ \bibinfo {pages} {123527} (\bibinfo {year} {2018}{\natexlab{a}})},\
  \Eprint {http://arxiv.org/abs/1809.06883} {arXiv:1809.06883 [astro-ph.CO]}
  \BibitemShut {NoStop}%
\bibitem [{\citenamefont {Di~Valentino}\ \emph
  {et~al.}(2021{\natexlab{a}})\citenamefont {Di~Valentino} \emph
  {et~al.}}]{DiValentino:2020vvd}%
  \BibitemOpen
  \bibfield  {author} {\bibinfo {author} {\bibfnamefont {Eleonora}\
  \bibnamefont {Di~Valentino}} \emph {et~al.},\ }\bibfield  {title} {\enquote
  {\bibinfo {title} {{Cosmology intertwined III: $f\sigma_8$ and $S_8$}},}\
  }\href {\doibase 10.1016/j.astropartphys.2021.102604} {\bibfield  {journal}
  {\bibinfo  {journal} {Astropart. Phys.}\ }\textbf {\bibinfo {volume} {131}},\
  \bibinfo {pages} {102604} (\bibinfo {year} {2021}{\natexlab{a}})},\ \Eprint
  {http://arxiv.org/abs/2008.11285} {arXiv:2008.11285 [astro-ph.CO]}
  \BibitemShut {NoStop}%
\bibitem [{\citenamefont {Di~Valentino}\ \emph
  {et~al.}(2021{\natexlab{b}})\citenamefont {Di~Valentino} \emph
  {et~al.}}]{DiValentino:2020zio}%
  \BibitemOpen
  \bibfield  {author} {\bibinfo {author} {\bibfnamefont {Eleonora}\
  \bibnamefont {Di~Valentino}} \emph {et~al.},\ }\bibfield  {title} {\enquote
  {\bibinfo {title} {{Snowmass2021 - Letter of interest cosmology intertwined
  II: The hubble constant tension}},}\ }\href {\doibase
  10.1016/j.astropartphys.2021.102605} {\bibfield  {journal} {\bibinfo
  {journal} {Astropart. Phys.}\ }\textbf {\bibinfo {volume} {131}},\ \bibinfo
  {pages} {102605} (\bibinfo {year} {2021}{\natexlab{b}})},\ \Eprint
  {http://arxiv.org/abs/2008.11284} {arXiv:2008.11284 [astro-ph.CO]}
  \BibitemShut {NoStop}%
\bibitem [{\citenamefont {Di~Valentino}\ \emph
  {et~al.}(2021{\natexlab{c}})\citenamefont {Di~Valentino}, \citenamefont
  {Mena}, \citenamefont {Pan}, \citenamefont {Visinelli}, \citenamefont {Yang},
  \citenamefont {Melchiorri}, \citenamefont {Mota}, \citenamefont {Riess},\
  and\ \citenamefont {Silk}}]{DiValentino:2021izs}%
  \BibitemOpen
  \bibfield  {author} {\bibinfo {author} {\bibfnamefont {Eleonora}\
  \bibnamefont {Di~Valentino}}, \bibinfo {author} {\bibfnamefont {Olga}\
  \bibnamefont {Mena}}, \bibinfo {author} {\bibfnamefont {Supriya}\
  \bibnamefont {Pan}}, \bibinfo {author} {\bibfnamefont {Luca}\ \bibnamefont
  {Visinelli}}, \bibinfo {author} {\bibfnamefont {Weiqiang}\ \bibnamefont
  {Yang}}, \bibinfo {author} {\bibfnamefont {Alessandro}\ \bibnamefont
  {Melchiorri}}, \bibinfo {author} {\bibfnamefont {David~F.}\ \bibnamefont
  {Mota}}, \bibinfo {author} {\bibfnamefont {Adam~G.}\ \bibnamefont {Riess}}, \
  and\ \bibinfo {author} {\bibfnamefont {Joseph}\ \bibnamefont {Silk}},\
  }\bibfield  {title} {\enquote {\bibinfo {title} {{In the realm of the Hubble
  tension\textemdash{}a review of solutions}},}\ }\href {\doibase
  10.1088/1361-6382/ac086d} {\bibfield  {journal} {\bibinfo  {journal} {Class.
  Quant. Grav.}\ }\textbf {\bibinfo {volume} {38}},\ \bibinfo {pages} {153001}
  (\bibinfo {year} {2021}{\natexlab{c}})},\ \Eprint
  {http://arxiv.org/abs/2103.01183} {arXiv:2103.01183 [astro-ph.CO]}
  \BibitemShut {NoStop}%
\bibitem [{\citenamefont {Barrow}\ and\ \citenamefont
  {Clifton}(2006)}]{Barrow:2006hia}%
  \BibitemOpen
  \bibfield  {author} {\bibinfo {author} {\bibfnamefont {John~D.}\ \bibnamefont
  {Barrow}}\ and\ \bibinfo {author} {\bibfnamefont {T.}~\bibnamefont
  {Clifton}},\ }\bibfield  {title} {\enquote {\bibinfo {title} {{Cosmologies
  with energy exchange}},}\ }\href {\doibase 10.1103/PhysRevD.73.103520}
  {\bibfield  {journal} {\bibinfo  {journal} {Phys. Rev. D}\ }\textbf {\bibinfo
  {volume} {73}},\ \bibinfo {pages} {103520} (\bibinfo {year} {2006})},\
  \Eprint {http://arxiv.org/abs/gr-qc/0604063} {arXiv:gr-qc/0604063}
  \BibitemShut {NoStop}%
\bibitem [{\citenamefont {Amendola}\ \emph {et~al.}(2007)\citenamefont
  {Amendola}, \citenamefont {Camargo~Campos},\ and\ \citenamefont
  {Rosenfeld}}]{Amendola:2006dg}%
  \BibitemOpen
  \bibfield  {author} {\bibinfo {author} {\bibfnamefont {Luca}\ \bibnamefont
  {Amendola}}, \bibinfo {author} {\bibfnamefont {Gabriela}\ \bibnamefont
  {Camargo~Campos}}, \ and\ \bibinfo {author} {\bibfnamefont {Rogerio}\
  \bibnamefont {Rosenfeld}},\ }\bibfield  {title} {\enquote {\bibinfo {title}
  {{Consequences of dark matter-dark energy interaction on cosmological
  parameters derived from SNIa data}},}\ }\href {\doibase
  10.1103/PhysRevD.75.083506} {\bibfield  {journal} {\bibinfo  {journal} {Phys.
  Rev. D}\ }\textbf {\bibinfo {volume} {75}},\ \bibinfo {pages} {083506}
  (\bibinfo {year} {2007})},\ \Eprint {http://arxiv.org/abs/astro-ph/0610806}
  {arXiv:astro-ph/0610806} \BibitemShut {NoStop}%
\bibitem [{\citenamefont {He}\ and\ \citenamefont {Wang}(2008)}]{He:2008tn}%
  \BibitemOpen
  \bibfield  {author} {\bibinfo {author} {\bibfnamefont {Jian-Hua}\
  \bibnamefont {He}}\ and\ \bibinfo {author} {\bibfnamefont {Bin}\ \bibnamefont
  {Wang}},\ }\bibfield  {title} {\enquote {\bibinfo {title} {{Effects of the
  interaction between dark energy and dark matter on cosmological
  parameters}},}\ }\href {\doibase 10.1088/1475-7516/2008/06/010} {\bibfield
  {journal} {\bibinfo  {journal} {JCAP}\ }\textbf {\bibinfo {volume} {06}},\
  \bibinfo {pages} {010} (\bibinfo {year} {2008})},\ \Eprint
  {http://arxiv.org/abs/0801.4233} {arXiv:0801.4233 [astro-ph]} \BibitemShut
  {NoStop}%
\bibitem [{\citenamefont {Basilakos}\ and\ \citenamefont
  {Plionis}(2009)}]{Basilakos:2008ae}%
  \BibitemOpen
  \bibfield  {author} {\bibinfo {author} {\bibfnamefont {Spyros}\ \bibnamefont
  {Basilakos}}\ and\ \bibinfo {author} {\bibfnamefont {M.}~\bibnamefont
  {Plionis}},\ }\bibfield  {title} {\enquote {\bibinfo {title} {{Is the
  Interacting Dark Matter Scenario an Alternative to Dark Energy ?}}}\ }\href
  {\doibase 10.1051/0004-6361/200912661} {\bibfield  {journal} {\bibinfo
  {journal} {Astron. Astrophys.}\ }\textbf {\bibinfo {volume} {507}},\ \bibinfo
  {pages} {47} (\bibinfo {year} {2009})},\ \Eprint
  {http://arxiv.org/abs/0807.4590} {arXiv:0807.4590 [astro-ph]} \BibitemShut
  {NoStop}%
\bibitem [{\citenamefont {Gavela}\ \emph {et~al.}(2009)\citenamefont {Gavela},
  \citenamefont {Hernandez}, \citenamefont {Lopez~Honorez}, \citenamefont
  {Mena},\ and\ \citenamefont {Rigolin}}]{Gavela:2009cy}%
  \BibitemOpen
  \bibfield  {author} {\bibinfo {author} {\bibfnamefont {M.~B.}\ \bibnamefont
  {Gavela}}, \bibinfo {author} {\bibfnamefont {D.}~\bibnamefont {Hernandez}},
  \bibinfo {author} {\bibfnamefont {L.}~\bibnamefont {Lopez~Honorez}}, \bibinfo
  {author} {\bibfnamefont {O.}~\bibnamefont {Mena}}, \ and\ \bibinfo {author}
  {\bibfnamefont {S.}~\bibnamefont {Rigolin}},\ }\bibfield  {title} {\enquote
  {\bibinfo {title} {{Dark coupling}},}\ }\href {\doibase
  10.1088/1475-7516/2009/07/034} {\bibfield  {journal} {\bibinfo  {journal}
  {JCAP}\ }\textbf {\bibinfo {volume} {07}},\ \bibinfo {pages} {034} (\bibinfo
  {year} {2009})},\ \bibinfo {note} {[Erratum: JCAP 05, E01 (2010)]},\ \Eprint
  {http://arxiv.org/abs/0901.1611} {arXiv:0901.1611 [astro-ph.CO]} \BibitemShut
  {NoStop}%
\bibitem [{\citenamefont {Chen}\ \emph {et~al.}(2014)\citenamefont {Chen},
  \citenamefont {Gong}, \citenamefont {Saridakis},\ and\ \citenamefont
  {Gong}}]{Chen:2011cy}%
  \BibitemOpen
  \bibfield  {author} {\bibinfo {author} {\bibfnamefont {Xi-ming}\ \bibnamefont
  {Chen}}, \bibinfo {author} {\bibfnamefont {Yungui}\ \bibnamefont {Gong}},
  \bibinfo {author} {\bibfnamefont {Emmanuel~N.}\ \bibnamefont {Saridakis}}, \
  and\ \bibinfo {author} {\bibfnamefont {Yungui}\ \bibnamefont {Gong}},\
  }\bibfield  {title} {\enquote {\bibinfo {title} {{Time-dependent interacting
  dark energy and transient acceleration}},}\ }\href {\doibase
  10.1007/s10773-013-1831-9} {\bibfield  {journal} {\bibinfo  {journal} {Int.
  J. Theor. Phys.}\ }\textbf {\bibinfo {volume} {53}},\ \bibinfo {pages}
  {469--481} (\bibinfo {year} {2014})},\ \Eprint
  {http://arxiv.org/abs/1111.6743} {arXiv:1111.6743 [astro-ph.CO]} \BibitemShut
  {NoStop}%
\bibitem [{\citenamefont {Pourtsidou}\ \emph {et~al.}(2013)\citenamefont
  {Pourtsidou}, \citenamefont {Skordis},\ and\ \citenamefont
  {Copeland}}]{Pourtsidou:2013nha}%
  \BibitemOpen
  \bibfield  {author} {\bibinfo {author} {\bibfnamefont {A.}~\bibnamefont
  {Pourtsidou}}, \bibinfo {author} {\bibfnamefont {C.}~\bibnamefont {Skordis}},
  \ and\ \bibinfo {author} {\bibfnamefont {E.~J.}\ \bibnamefont {Copeland}},\
  }\bibfield  {title} {\enquote {\bibinfo {title} {{Models of dark matter
  coupled to dark energy}},}\ }\href {\doibase 10.1103/PhysRevD.88.083505}
  {\bibfield  {journal} {\bibinfo  {journal} {Phys. Rev. D}\ }\textbf {\bibinfo
  {volume} {88}},\ \bibinfo {pages} {083505} (\bibinfo {year} {2013})},\
  \Eprint {http://arxiv.org/abs/1307.0458} {arXiv:1307.0458 [astro-ph.CO]}
  \BibitemShut {NoStop}%
\bibitem [{\citenamefont {Yang}\ and\ \citenamefont {Xu}(2014)}]{Yang:2014hea}%
  \BibitemOpen
  \bibfield  {author} {\bibinfo {author} {\bibfnamefont {Weiqiang}\
  \bibnamefont {Yang}}\ and\ \bibinfo {author} {\bibfnamefont {Lixin}\
  \bibnamefont {Xu}},\ }\bibfield  {title} {\enquote {\bibinfo {title}
  {{Coupled dark energy with perturbed Hubble expansion rate}},}\ }\href
  {\doibase 10.1103/PhysRevD.90.083532} {\bibfield  {journal} {\bibinfo
  {journal} {Phys. Rev. D}\ }\textbf {\bibinfo {volume} {90}},\ \bibinfo
  {pages} {083532} (\bibinfo {year} {2014})},\ \Eprint
  {http://arxiv.org/abs/1409.5533} {arXiv:1409.5533 [astro-ph.CO]} \BibitemShut
  {NoStop}%
\bibitem [{\citenamefont {Nunes}\ and\ \citenamefont
  {Barboza}(2014)}]{Nunes:2014qoa}%
  \BibitemOpen
  \bibfield  {author} {\bibinfo {author} {\bibfnamefont {Rafael~C.}\
  \bibnamefont {Nunes}}\ and\ \bibinfo {author} {\bibfnamefont {Edesio~M.}\
  \bibnamefont {Barboza}},\ }\bibfield  {title} {\enquote {\bibinfo {title}
  {{Dark matter-dark energy interaction for a time-dependent EoS parameter}},}\
  }\href {\doibase 10.1007/s10714-014-1820-1} {\bibfield  {journal} {\bibinfo
  {journal} {Gen. Rel. Grav.}\ }\textbf {\bibinfo {volume} {46}},\ \bibinfo
  {pages} {1820} (\bibinfo {year} {2014})},\ \Eprint
  {http://arxiv.org/abs/1404.1620} {arXiv:1404.1620 [astro-ph.CO]} \BibitemShut
  {NoStop}%
\bibitem [{\citenamefont {Faraoni}\ \emph {et~al.}(2014)\citenamefont
  {Faraoni}, \citenamefont {Dent},\ and\ \citenamefont
  {Saridakis}}]{Faraoni:2014vra}%
  \BibitemOpen
  \bibfield  {author} {\bibinfo {author} {\bibfnamefont {Valerio}\ \bibnamefont
  {Faraoni}}, \bibinfo {author} {\bibfnamefont {James~B.}\ \bibnamefont
  {Dent}}, \ and\ \bibinfo {author} {\bibfnamefont {Emmanuel~N.}\ \bibnamefont
  {Saridakis}},\ }\bibfield  {title} {\enquote {\bibinfo {title}
  {{Covariantizing the interaction between dark energy and dark matter}},}\
  }\href {\doibase 10.1103/PhysRevD.90.063510} {\bibfield  {journal} {\bibinfo
  {journal} {Phys. Rev. D}\ }\textbf {\bibinfo {volume} {90}},\ \bibinfo
  {pages} {063510} (\bibinfo {year} {2014})},\ \Eprint
  {http://arxiv.org/abs/1405.7288} {arXiv:1405.7288 [gr-qc]} \BibitemShut
  {NoStop}%
\bibitem [{\citenamefont {Salvatelli}\ \emph {et~al.}(2014)\citenamefont
  {Salvatelli}, \citenamefont {Said}, \citenamefont {Bruni}, \citenamefont
  {Melchiorri},\ and\ \citenamefont {Wands}}]{Salvatelli:2014zta}%
  \BibitemOpen
  \bibfield  {author} {\bibinfo {author} {\bibfnamefont {Valentina}\
  \bibnamefont {Salvatelli}}, \bibinfo {author} {\bibfnamefont {Najla}\
  \bibnamefont {Said}}, \bibinfo {author} {\bibfnamefont {Marco}\ \bibnamefont
  {Bruni}}, \bibinfo {author} {\bibfnamefont {Alessandro}\ \bibnamefont
  {Melchiorri}}, \ and\ \bibinfo {author} {\bibfnamefont {David}\ \bibnamefont
  {Wands}},\ }\bibfield  {title} {\enquote {\bibinfo {title} {{Indications of a
  late-time interaction in the dark sector}},}\ }\href {\doibase
  10.1103/PhysRevLett.113.181301} {\bibfield  {journal} {\bibinfo  {journal}
  {Phys. Rev. Lett.}\ }\textbf {\bibinfo {volume} {113}},\ \bibinfo {pages}
  {181301} (\bibinfo {year} {2014})},\ \Eprint {http://arxiv.org/abs/1406.7297}
  {arXiv:1406.7297 [astro-ph.CO]} \BibitemShut {NoStop}%
\bibitem [{\citenamefont {Pan}\ \emph {et~al.}(2015)\citenamefont {Pan},
  \citenamefont {Bhattacharya},\ and\ \citenamefont
  {Chakraborty}}]{Pan:2012ki}%
  \BibitemOpen
  \bibfield  {author} {\bibinfo {author} {\bibfnamefont {Supriya}\ \bibnamefont
  {Pan}}, \bibinfo {author} {\bibfnamefont {Subhra}\ \bibnamefont
  {Bhattacharya}}, \ and\ \bibinfo {author} {\bibfnamefont {Subenoy}\
  \bibnamefont {Chakraborty}},\ }\bibfield  {title} {\enquote {\bibinfo {title}
  {{An analytic model for interacting dark energy and its observational
  constraints}},}\ }\href {\doibase 10.1093/mnras/stv1495} {\bibfield
  {journal} {\bibinfo  {journal} {Mon. Not. Roy. Astron. Soc.}\ }\textbf
  {\bibinfo {volume} {452}},\ \bibinfo {pages} {3038--3046} (\bibinfo {year}
  {2015})},\ \Eprint {http://arxiv.org/abs/1210.0396} {arXiv:1210.0396 [gr-qc]}
  \BibitemShut {NoStop}%
\bibitem [{\citenamefont {Boehmer}\ \emph
  {et~al.}(2015{\natexlab{a}})\citenamefont {Boehmer}, \citenamefont
  {Tamanini},\ and\ \citenamefont {Wright}}]{Boehmer:2015kta}%
  \BibitemOpen
  \bibfield  {author} {\bibinfo {author} {\bibfnamefont {Christian~G.}\
  \bibnamefont {Boehmer}}, \bibinfo {author} {\bibfnamefont {Nicola}\
  \bibnamefont {Tamanini}}, \ and\ \bibinfo {author} {\bibfnamefont {Matthew}\
  \bibnamefont {Wright}},\ }\bibfield  {title} {\enquote {\bibinfo {title}
  {{Interacting quintessence from a variational approach Part I: algebraic
  couplings}},}\ }\href {\doibase 10.1103/PhysRevD.91.123002} {\bibfield
  {journal} {\bibinfo  {journal} {Phys. Rev. D}\ }\textbf {\bibinfo {volume}
  {91}},\ \bibinfo {pages} {123002} (\bibinfo {year} {2015}{\natexlab{a}})},\
  \Eprint {http://arxiv.org/abs/1501.06540} {arXiv:1501.06540 [gr-qc]}
  \BibitemShut {NoStop}%
\bibitem [{\citenamefont {Nunes}\ \emph {et~al.}(2016)\citenamefont {Nunes},
  \citenamefont {Pan},\ and\ \citenamefont {Saridakis}}]{Nunes:2016dlj}%
  \BibitemOpen
  \bibfield  {author} {\bibinfo {author} {\bibfnamefont {Rafael~C.}\
  \bibnamefont {Nunes}}, \bibinfo {author} {\bibfnamefont {Supriya}\
  \bibnamefont {Pan}}, \ and\ \bibinfo {author} {\bibfnamefont {Emmanuel~N.}\
  \bibnamefont {Saridakis}},\ }\bibfield  {title} {\enquote {\bibinfo {title}
  {{New constraints on interacting dark energy from cosmic chronometers}},}\
  }\href {\doibase 10.1103/PhysRevD.94.023508} {\bibfield  {journal} {\bibinfo
  {journal} {Phys. Rev. D}\ }\textbf {\bibinfo {volume} {94}},\ \bibinfo
  {pages} {023508} (\bibinfo {year} {2016})},\ \Eprint
  {http://arxiv.org/abs/1605.01712} {arXiv:1605.01712 [astro-ph.CO]}
  \BibitemShut {NoStop}%
\bibitem [{\citenamefont {Boehmer}\ \emph
  {et~al.}(2015{\natexlab{b}})\citenamefont {Boehmer}, \citenamefont
  {Tamanini},\ and\ \citenamefont {Wright}}]{Boehmer:2015sha}%
  \BibitemOpen
  \bibfield  {author} {\bibinfo {author} {\bibfnamefont {Christian~G.}\
  \bibnamefont {Boehmer}}, \bibinfo {author} {\bibfnamefont {Nicola}\
  \bibnamefont {Tamanini}}, \ and\ \bibinfo {author} {\bibfnamefont {Matthew}\
  \bibnamefont {Wright}},\ }\bibfield  {title} {\enquote {\bibinfo {title}
  {{Interacting quintessence from a variational approach Part II: derivative
  couplings}},}\ }\href {\doibase 10.1103/PhysRevD.91.123003} {\bibfield
  {journal} {\bibinfo  {journal} {Phys. Rev. D}\ }\textbf {\bibinfo {volume}
  {91}},\ \bibinfo {pages} {123003} (\bibinfo {year} {2015}{\natexlab{b}})},\
  \Eprint {http://arxiv.org/abs/1502.04030} {arXiv:1502.04030 [gr-qc]}
  \BibitemShut {NoStop}%
\bibitem [{\citenamefont {Mukherjee}\ and\ \citenamefont
  {Banerjee}(2017)}]{Mukherjee:2016shl}%
  \BibitemOpen
  \bibfield  {author} {\bibinfo {author} {\bibfnamefont {Ankan}\ \bibnamefont
  {Mukherjee}}\ and\ \bibinfo {author} {\bibfnamefont {Narayan}\ \bibnamefont
  {Banerjee}},\ }\bibfield  {title} {\enquote {\bibinfo {title} {{In search of
  the dark matter dark energy interaction: a kinematic approach}},}\ }\href
  {\doibase 10.1088/1361-6382/aa54c8} {\bibfield  {journal} {\bibinfo
  {journal} {Class. Quant. Grav.}\ }\textbf {\bibinfo {volume} {34}},\ \bibinfo
  {pages} {035016} (\bibinfo {year} {2017})},\ \Eprint
  {http://arxiv.org/abs/1610.04419} {arXiv:1610.04419 [astro-ph.CO]}
  \BibitemShut {NoStop}%
\bibitem [{\citenamefont {Cai}\ \emph {et~al.}(2017)\citenamefont {Cai},
  \citenamefont {Tamanini},\ and\ \citenamefont {Yang}}]{Cai:2017yww}%
  \BibitemOpen
  \bibfield  {author} {\bibinfo {author} {\bibfnamefont {Rong-Gen}\
  \bibnamefont {Cai}}, \bibinfo {author} {\bibfnamefont {Nicola}\ \bibnamefont
  {Tamanini}}, \ and\ \bibinfo {author} {\bibfnamefont {Tao}\ \bibnamefont
  {Yang}},\ }\bibfield  {title} {\enquote {\bibinfo {title} {{Reconstructing
  the dark sector interaction with LISA}},}\ }\href {\doibase
  10.1088/1475-7516/2017/05/031} {\bibfield  {journal} {\bibinfo  {journal}
  {JCAP}\ }\textbf {\bibinfo {volume} {05}},\ \bibinfo {pages} {031} (\bibinfo
  {year} {2017})},\ \Eprint {http://arxiv.org/abs/1703.07323} {arXiv:1703.07323
  [astro-ph.CO]} \BibitemShut {NoStop}%
\bibitem [{\citenamefont {Yang}\ \emph
  {et~al.}(2018{\natexlab{b}})\citenamefont {Yang}, \citenamefont {Pan},\ and\
  \citenamefont {Barrow}}]{Yang:2017zjs}%
  \BibitemOpen
  \bibfield  {author} {\bibinfo {author} {\bibfnamefont {Weiqiang}\
  \bibnamefont {Yang}}, \bibinfo {author} {\bibfnamefont {Supriya}\
  \bibnamefont {Pan}}, \ and\ \bibinfo {author} {\bibfnamefont {John~D.}\
  \bibnamefont {Barrow}},\ }\bibfield  {title} {\enquote {\bibinfo {title}
  {{Large-scale Stability and Astronomical Constraints for Coupled Dark-Energy
  Models}},}\ }\href {\doibase 10.1103/PhysRevD.97.043529} {\bibfield
  {journal} {\bibinfo  {journal} {Phys. Rev. D}\ }\textbf {\bibinfo {volume}
  {97}},\ \bibinfo {pages} {043529} (\bibinfo {year} {2018}{\natexlab{b}})},\
  \Eprint {http://arxiv.org/abs/1706.04953} {arXiv:1706.04953 [astro-ph.CO]}
  \BibitemShut {NoStop}%
\bibitem [{\citenamefont {Santos}\ \emph {et~al.}(2017)\citenamefont {Santos},
  \citenamefont {Zhao}, \citenamefont {Ferreira},\ and\ \citenamefont
  {Quintin}}]{Santos:2017bqm}%
  \BibitemOpen
  \bibfield  {author} {\bibinfo {author} {\bibfnamefont {Larissa}\ \bibnamefont
  {Santos}}, \bibinfo {author} {\bibfnamefont {Wen}\ \bibnamefont {Zhao}},
  \bibinfo {author} {\bibfnamefont {Elisa G.~M.}\ \bibnamefont {Ferreira}}, \
  and\ \bibinfo {author} {\bibfnamefont {Jerome}\ \bibnamefont {Quintin}},\
  }\bibfield  {title} {\enquote {\bibinfo {title} {{Constraining interacting
  dark energy with CMB and BAO future surveys}},}\ }\href {\doibase
  10.1103/PhysRevD.96.103529} {\bibfield  {journal} {\bibinfo  {journal} {Phys.
  Rev. D}\ }\textbf {\bibinfo {volume} {96}},\ \bibinfo {pages} {103529}
  (\bibinfo {year} {2017})},\ \Eprint {http://arxiv.org/abs/1707.06827}
  {arXiv:1707.06827 [astro-ph.CO]} \BibitemShut {NoStop}%
\bibitem [{\citenamefont {Pan}\ \emph {et~al.}(2018)\citenamefont {Pan},
  \citenamefont {Mukherjee},\ and\ \citenamefont {Banerjee}}]{Pan:2017ent}%
  \BibitemOpen
  \bibfield  {author} {\bibinfo {author} {\bibfnamefont {Supriya}\ \bibnamefont
  {Pan}}, \bibinfo {author} {\bibfnamefont {Ankan}\ \bibnamefont {Mukherjee}},
  \ and\ \bibinfo {author} {\bibfnamefont {Narayan}\ \bibnamefont {Banerjee}},\
  }\bibfield  {title} {\enquote {\bibinfo {title} {{Astronomical bounds on a
  cosmological model allowing a general interaction in the dark sector}},}\
  }\href {\doibase 10.1093/mnras/sty755} {\bibfield  {journal} {\bibinfo
  {journal} {Mon. Not. Roy. Astron. Soc.}\ }\textbf {\bibinfo {volume} {477}},\
  \bibinfo {pages} {1189--1205} (\bibinfo {year} {2018})},\ \Eprint
  {http://arxiv.org/abs/1710.03725} {arXiv:1710.03725 [astro-ph.CO]}
  \BibitemShut {NoStop}%
\bibitem [{\citenamefont {Grandon}\ and\ \citenamefont
  {Cardenas}(2018)}]{Grandon:2018uoe}%
  \BibitemOpen
  \bibfield  {author} {\bibinfo {author} {\bibfnamefont {Daniela}\ \bibnamefont
  {Grandon}}\ and\ \bibinfo {author} {\bibfnamefont {Victor~H.}\ \bibnamefont
  {Cardenas}},\ }\bibfield  {title} {\enquote {\bibinfo {title} {{Exploring
  evidence of interaction between dark energy and dark matter}},}\ }\href
  {\doibase 10.1007/s10714-019-2526-1} {\  (\bibinfo {year} {2018}),\
  10.1007/s10714-019-2526-1},\ \Eprint {http://arxiv.org/abs/1804.03296}
  {arXiv:1804.03296 [astro-ph.CO]} \BibitemShut {NoStop}%
\bibitem [{\citenamefont {von Marttens}\ \emph {et~al.}(2019)\citenamefont {von
  Marttens}, \citenamefont {Casarini}, \citenamefont {Mota},\ and\
  \citenamefont {Zimdahl}}]{vonMarttens:2018iav}%
  \BibitemOpen
  \bibfield  {author} {\bibinfo {author} {\bibfnamefont {R.}~\bibnamefont {von
  Marttens}}, \bibinfo {author} {\bibfnamefont {L.}~\bibnamefont {Casarini}},
  \bibinfo {author} {\bibfnamefont {D.~F.}\ \bibnamefont {Mota}}, \ and\
  \bibinfo {author} {\bibfnamefont {W.}~\bibnamefont {Zimdahl}},\ }\bibfield
  {title} {\enquote {\bibinfo {title} {{Cosmological constraints on
  parametrized interacting dark energy}},}\ }\href {\doibase
  10.1016/j.dark.2018.10.007} {\bibfield  {journal} {\bibinfo  {journal} {Phys.
  Dark Univ.}\ }\textbf {\bibinfo {volume} {23}},\ \bibinfo {pages} {100248}
  (\bibinfo {year} {2019})},\ \Eprint {http://arxiv.org/abs/1807.11380}
  {arXiv:1807.11380 [astro-ph.CO]} \BibitemShut {NoStop}%
\bibitem [{\citenamefont {Bonici}\ and\ \citenamefont
  {Maggiore}(2019)}]{Bonici:2018qli}%
  \BibitemOpen
  \bibfield  {author} {\bibinfo {author} {\bibfnamefont {Marco}\ \bibnamefont
  {Bonici}}\ and\ \bibinfo {author} {\bibfnamefont {Nicola}\ \bibnamefont
  {Maggiore}},\ }\bibfield  {title} {\enquote {\bibinfo {title} {{Constraints
  on interacting dynamical dark energy and a new test for $\Lambda $ CDM}},}\
  }\href {\doibase 10.1140/epjc/s10052-019-7198-1} {\bibfield  {journal}
  {\bibinfo  {journal} {Eur. Phys. J. C}\ }\textbf {\bibinfo {volume} {79}},\
  \bibinfo {pages} {672} (\bibinfo {year} {2019})},\ \Eprint
  {http://arxiv.org/abs/1812.11176} {arXiv:1812.11176 [gr-qc]} \BibitemShut
  {NoStop}%
\bibitem [{\citenamefont {Li}\ \emph {et~al.}(2020)\citenamefont {Li},
  \citenamefont {Ren}, \citenamefont {Khurshudyan},\ and\ \citenamefont
  {Cai}}]{Li:2019loh}%
  \BibitemOpen
  \bibfield  {author} {\bibinfo {author} {\bibfnamefont {Chunlong}\
  \bibnamefont {Li}}, \bibinfo {author} {\bibfnamefont {Xin}\ \bibnamefont
  {Ren}}, \bibinfo {author} {\bibfnamefont {Martiros}\ \bibnamefont
  {Khurshudyan}}, \ and\ \bibinfo {author} {\bibfnamefont {Yi-Fu}\ \bibnamefont
  {Cai}},\ }\bibfield  {title} {\enquote {\bibinfo {title} {{Implications of
  the possible 21-cm line excess at cosmic dawn on dynamics of interacting dark
  energy}},}\ }\href {\doibase 10.1016/j.physletb.2019.135141} {\bibfield
  {journal} {\bibinfo  {journal} {Phys. Lett. B}\ }\textbf {\bibinfo {volume}
  {801}},\ \bibinfo {pages} {135141} (\bibinfo {year} {2020})},\ \Eprint
  {http://arxiv.org/abs/1904.02458} {arXiv:1904.02458 [astro-ph.CO]}
  \BibitemShut {NoStop}%
\bibitem [{\citenamefont {Yang}\ \emph {et~al.}(2020)\citenamefont {Yang},
  \citenamefont {Pan}, \citenamefont {Di~Valentino}, \citenamefont {Wang},\
  and\ \citenamefont {Wang}}]{Yang:2019bpr}%
  \BibitemOpen
  \bibfield  {author} {\bibinfo {author} {\bibfnamefont {Weiqiang}\
  \bibnamefont {Yang}}, \bibinfo {author} {\bibfnamefont {Supriya}\
  \bibnamefont {Pan}}, \bibinfo {author} {\bibfnamefont {Eleonora}\
  \bibnamefont {Di~Valentino}}, \bibinfo {author} {\bibfnamefont {Bin}\
  \bibnamefont {Wang}}, \ and\ \bibinfo {author} {\bibfnamefont {Anzhong}\
  \bibnamefont {Wang}},\ }\bibfield  {title} {\enquote {\bibinfo {title}
  {{Forecasting interacting vacuum-energy models using gravitational waves}},}\
  }\href {\doibase 10.1088/1475-7516/2020/05/050} {\bibfield  {journal}
  {\bibinfo  {journal} {JCAP}\ }\textbf {\bibinfo {volume} {05}},\ \bibinfo
  {pages} {050} (\bibinfo {year} {2020})},\ \Eprint
  {http://arxiv.org/abs/1904.11980} {arXiv:1904.11980 [astro-ph.CO]}
  \BibitemShut {NoStop}%
\bibitem [{\citenamefont {Mamon}\ \emph {et~al.}(2021)\citenamefont {Mamon},
  \citenamefont {Paliathanasis},\ and\ \citenamefont {Saha}}]{Mamon:2020spa}%
  \BibitemOpen
  \bibfield  {author} {\bibinfo {author} {\bibfnamefont {Abdulla~Al}\
  \bibnamefont {Mamon}}, \bibinfo {author} {\bibfnamefont {Andronikos}\
  \bibnamefont {Paliathanasis}}, \ and\ \bibinfo {author} {\bibfnamefont
  {Subhajit}\ \bibnamefont {Saha}},\ }\bibfield  {title} {\enquote {\bibinfo
  {title} {{Dynamics of an Interacting Barrow Holographic Dark Energy Model and
  its Thermodynamic Implications}},}\ }\href {\doibase
  10.1140/epjp/s13360-021-01130-7} {\bibfield  {journal} {\bibinfo  {journal}
  {Eur. Phys. J. Plus}\ }\textbf {\bibinfo {volume} {136}},\ \bibinfo {pages}
  {134} (\bibinfo {year} {2021})},\ \Eprint {http://arxiv.org/abs/2007.16020}
  {arXiv:2007.16020 [gr-qc]} \BibitemShut {NoStop}%
\bibitem [{\citenamefont {Beltr\'an~Jim\'enez}\ \emph
  {et~al.}(2021)\citenamefont {Beltr\'an~Jim\'enez}, \citenamefont {Bettoni},
  \citenamefont {Figueruelo}, \citenamefont {Teppa~Pannia},\ and\ \citenamefont
  {Tsujikawa}}]{BeltranJimenez:2020qdu}%
  \BibitemOpen
  \bibfield  {author} {\bibinfo {author} {\bibfnamefont {Jose}\ \bibnamefont
  {Beltr\'an~Jim\'enez}}, \bibinfo {author} {\bibfnamefont {Dario}\
  \bibnamefont {Bettoni}}, \bibinfo {author} {\bibfnamefont {David}\
  \bibnamefont {Figueruelo}}, \bibinfo {author} {\bibfnamefont {Florencia~A.}\
  \bibnamefont {Teppa~Pannia}}, \ and\ \bibinfo {author} {\bibfnamefont
  {Shinji}\ \bibnamefont {Tsujikawa}},\ }\bibfield  {title} {\enquote {\bibinfo
  {title} {{Velocity-dependent interacting dark energy and dark matter with a
  Lagrangian description of perfect fluids}},}\ }\href {\doibase
  10.1088/1475-7516/2021/03/085} {\bibfield  {journal} {\bibinfo  {journal}
  {JCAP}\ }\textbf {\bibinfo {volume} {03}},\ \bibinfo {pages} {085} (\bibinfo
  {year} {2021})},\ \Eprint {http://arxiv.org/abs/2012.12204} {arXiv:2012.12204
  [astro-ph.CO]} \BibitemShut {NoStop}%
\bibitem [{\citenamefont {Lucca}(2021)}]{Lucca:2021eqy}%
  \BibitemOpen
  \bibfield  {author} {\bibinfo {author} {\bibfnamefont {Matteo}\ \bibnamefont
  {Lucca}},\ }\bibfield  {title} {\enquote {\bibinfo {title}
  {{Multi-interacting dark energy and its cosmological implications}},}\ }\href
  {\doibase 10.1103/PhysRevD.104.083510} {\bibfield  {journal} {\bibinfo
  {journal} {Phys. Rev. D}\ }\textbf {\bibinfo {volume} {104}},\ \bibinfo
  {pages} {083510} (\bibinfo {year} {2021})},\ \Eprint
  {http://arxiv.org/abs/2106.15196} {arXiv:2106.15196 [astro-ph.CO]}
  \BibitemShut {NoStop}%
\bibitem [{\citenamefont {Konitopoulos}\ \emph {et~al.}(2021)\citenamefont
  {Konitopoulos}, \citenamefont {Saridakis}, \citenamefont {Stavrinos},\ and\
  \citenamefont {Triantafyllopoulos}}]{Konitopoulos:2021eav}%
  \BibitemOpen
  \bibfield  {author} {\bibinfo {author} {\bibfnamefont {Spyros}\ \bibnamefont
  {Konitopoulos}}, \bibinfo {author} {\bibfnamefont {Emmanuel~N.}\ \bibnamefont
  {Saridakis}}, \bibinfo {author} {\bibfnamefont {P.~C.}\ \bibnamefont
  {Stavrinos}}, \ and\ \bibinfo {author} {\bibfnamefont {A.}~\bibnamefont
  {Triantafyllopoulos}},\ }\bibfield  {title} {\enquote {\bibinfo {title}
  {{Dark gravitational sectors on a generalized scalar-tensor vector bundle
  model and cosmological applications}},}\ }\href {\doibase
  10.1103/PhysRevD.104.064018} {\bibfield  {journal} {\bibinfo  {journal}
  {Phys. Rev. D}\ }\textbf {\bibinfo {volume} {104}},\ \bibinfo {pages}
  {064018} (\bibinfo {year} {2021})},\ \Eprint
  {http://arxiv.org/abs/2104.08024} {arXiv:2104.08024 [gr-qc]} \BibitemShut
  {NoStop}%
\bibitem [{\citenamefont {Bolotin}\ \emph {et~al.}(2014)\citenamefont
  {Bolotin}, \citenamefont {Kostenko}, \citenamefont {Lemets},\ and\
  \citenamefont {Yerokhin}}]{Bolotin:2013jpa}%
  \BibitemOpen
  \bibfield  {author} {\bibinfo {author} {\bibfnamefont {Yu.~L.}\ \bibnamefont
  {Bolotin}}, \bibinfo {author} {\bibfnamefont {A.}~\bibnamefont {Kostenko}},
  \bibinfo {author} {\bibfnamefont {O.~A.}\ \bibnamefont {Lemets}}, \ and\
  \bibinfo {author} {\bibfnamefont {D.~A.}\ \bibnamefont {Yerokhin}},\
  }\bibfield  {title} {\enquote {\bibinfo {title} {{Cosmological Evolution With
  Interaction Between Dark Energy And Dark Matter}},}\ }\href {\doibase
  10.1142/S0218271815300074} {\bibfield  {journal} {\bibinfo  {journal} {Int.
  J. Mod. Phys. D}\ }\textbf {\bibinfo {volume} {24}},\ \bibinfo {pages}
  {1530007} (\bibinfo {year} {2014})},\ \Eprint
  {http://arxiv.org/abs/1310.0085} {arXiv:1310.0085 [astro-ph.CO]} \BibitemShut
  {NoStop}%
\bibitem [{\citenamefont {Wang}\ \emph {et~al.}(2016)\citenamefont {Wang},
  \citenamefont {Abdalla}, \citenamefont {Atrio-Barandela},\ and\ \citenamefont
  {Pavon}}]{Wang:2016lxa}%
  \BibitemOpen
  \bibfield  {author} {\bibinfo {author} {\bibfnamefont {B.}~\bibnamefont
  {Wang}}, \bibinfo {author} {\bibfnamefont {E.}~\bibnamefont {Abdalla}},
  \bibinfo {author} {\bibfnamefont {F.}~\bibnamefont {Atrio-Barandela}}, \ and\
  \bibinfo {author} {\bibfnamefont {D.}~\bibnamefont {Pavon}},\ }\bibfield
  {title} {\enquote {\bibinfo {title} {{Dark Matter and Dark Energy
  Interactions: Theoretical Challenges, Cosmological Implications and
  Observational Signatures}},}\ }\href {\doibase 10.1088/0034-4885/79/9/096901}
  {\bibfield  {journal} {\bibinfo  {journal} {Rept. Prog. Phys.}\ }\textbf
  {\bibinfo {volume} {79}},\ \bibinfo {pages} {096901} (\bibinfo {year}
  {2016})},\ \Eprint {http://arxiv.org/abs/1603.08299} {arXiv:1603.08299
  [astro-ph.CO]} \BibitemShut {NoStop}%
\bibitem [{\citenamefont {Abdalla}\ \emph {et~al.}(2009)\citenamefont
  {Abdalla}, \citenamefont {Abramo}, \citenamefont {Sodre},\ and\ \citenamefont
  {Wang}}]{Abdalla:2007rd}%
  \BibitemOpen
  \bibfield  {author} {\bibinfo {author} {\bibfnamefont {E.}~\bibnamefont
  {Abdalla}}, \bibinfo {author} {\bibfnamefont {L.~Raul~W.}\ \bibnamefont
  {Abramo}}, \bibinfo {author} {\bibfnamefont {L.}~\bibnamefont {Sodre},
  \bibfnamefont {Jr.}}, \ and\ \bibinfo {author} {\bibfnamefont
  {B.}~\bibnamefont {Wang}},\ }\bibfield  {title} {\enquote {\bibinfo {title}
  {{Signature of the interaction between dark energy and dark matter in galaxy
  clusters}},}\ }\href {\doibase 10.1016/j.physletb.2009.02.008} {\bibfield
  {journal} {\bibinfo  {journal} {Phys. Lett. B}\ }\textbf {\bibinfo {volume}
  {673}},\ \bibinfo {pages} {107--110} (\bibinfo {year} {2009})},\ \Eprint
  {http://arxiv.org/abs/0710.1198} {arXiv:0710.1198 [astro-ph]} \BibitemShut
  {NoStop}%
\bibitem [{\citenamefont {He}\ \emph {et~al.}(2011)\citenamefont {He},
  \citenamefont {Wang},\ and\ \citenamefont {Abdalla}}]{He:2010im}%
  \BibitemOpen
  \bibfield  {author} {\bibinfo {author} {\bibfnamefont {Jian-Hua}\
  \bibnamefont {He}}, \bibinfo {author} {\bibfnamefont {Bin}\ \bibnamefont
  {Wang}}, \ and\ \bibinfo {author} {\bibfnamefont {Elcio}\ \bibnamefont
  {Abdalla}},\ }\bibfield  {title} {\enquote {\bibinfo {title} {{Testing the
  interaction between dark energy and dark matter via latest observations}},}\
  }\href {\doibase 10.1103/PhysRevD.83.063515} {\bibfield  {journal} {\bibinfo
  {journal} {Phys. Rev. D}\ }\textbf {\bibinfo {volume} {83}},\ \bibinfo
  {pages} {063515} (\bibinfo {year} {2011})},\ \Eprint
  {http://arxiv.org/abs/1012.3904} {arXiv:1012.3904 [astro-ph.CO]} \BibitemShut
  {NoStop}%
\bibitem [{\citenamefont {Solano}\ and\ \citenamefont
  {Nucamendi}(2012)}]{Solano:2011ie}%
  \BibitemOpen
  \bibfield  {author} {\bibinfo {author} {\bibfnamefont {Freddy~Cueva}\
  \bibnamefont {Solano}}\ and\ \bibinfo {author} {\bibfnamefont {Ulises}\
  \bibnamefont {Nucamendi}},\ }\bibfield  {title} {\enquote {\bibinfo {title}
  {{Reconstruction of the interaction term between dark matter and dark energy
  using SNe Ia}},}\ }\href {\doibase 10.1088/1475-7516/2012/04/011} {\bibfield
  {journal} {\bibinfo  {journal} {JCAP}\ }\textbf {\bibinfo {volume} {04}},\
  \bibinfo {pages} {011} (\bibinfo {year} {2012})},\ \Eprint
  {http://arxiv.org/abs/1109.1303} {arXiv:1109.1303 [astro-ph.CO]} \BibitemShut
  {NoStop}%
\bibitem [{\citenamefont {Cao}\ and\ \citenamefont
  {Liang}(2013)}]{doi:10.1142/S021827181350082X}%
  \BibitemOpen
  \bibfield  {author} {\bibinfo {author} {\bibfnamefont {Shuo}\ \bibnamefont
  {Cao}}\ and\ \bibinfo {author} {\bibfnamefont {Nan}\ \bibnamefont {Liang}},\
  }\bibfield  {title} {\enquote {\bibinfo {title} {Interaction between dark
  energy and dark matter: Observational constraints from ohd, bao, cmb and sne
  ia},}\ }\href {\doibase 10.1142/S021827181350082X} {\bibfield  {journal}
  {\bibinfo  {journal} {International Journal of Modern Physics D}\ }\textbf
  {\bibinfo {volume} {22}},\ \bibinfo {pages} {1350082} (\bibinfo {year}
  {2013})},\ \Eprint
  {http://arxiv.org/abs/https://doi.org/10.1142/S021827181350082X}
  {https://doi.org/10.1142/S021827181350082X} \BibitemShut {NoStop}%
\bibitem [{\citenamefont {Costa}\ \emph {et~al.}(2014)\citenamefont {Costa},
  \citenamefont {Xu}, \citenamefont {Wang}, \citenamefont {Ferreira},\ and\
  \citenamefont {Abdalla}}]{Costa:2013sva}%
  \BibitemOpen
  \bibfield  {author} {\bibinfo {author} {\bibfnamefont {Andr\'e~A.}\
  \bibnamefont {Costa}}, \bibinfo {author} {\bibfnamefont {Xiao-Dong}\
  \bibnamefont {Xu}}, \bibinfo {author} {\bibfnamefont {Bin}\ \bibnamefont
  {Wang}}, \bibinfo {author} {\bibfnamefont {Elisa G.~M.}\ \bibnamefont
  {Ferreira}}, \ and\ \bibinfo {author} {\bibfnamefont {E.}~\bibnamefont
  {Abdalla}},\ }\bibfield  {title} {\enquote {\bibinfo {title} {{Testing the
  Interaction between Dark Energy and Dark Matter with Planck Data}},}\ }\href
  {\doibase 10.1103/PhysRevD.89.103531} {\bibfield  {journal} {\bibinfo
  {journal} {Phys. Rev. D}\ }\textbf {\bibinfo {volume} {89}},\ \bibinfo
  {pages} {103531} (\bibinfo {year} {2014})},\ \Eprint
  {http://arxiv.org/abs/1311.7380} {arXiv:1311.7380 [astro-ph.CO]} \BibitemShut
  {NoStop}%
\bibitem [{\citenamefont {Pan}\ \emph {et~al.}(2019{\natexlab{a}})\citenamefont
  {Pan}, \citenamefont {Yang}, \citenamefont {Singha},\ and\ \citenamefont
  {Saridakis}}]{Pan:2019jqh}%
  \BibitemOpen
  \bibfield  {author} {\bibinfo {author} {\bibfnamefont {Supriya}\ \bibnamefont
  {Pan}}, \bibinfo {author} {\bibfnamefont {Weiqiang}\ \bibnamefont {Yang}},
  \bibinfo {author} {\bibfnamefont {Chiranjeeb}\ \bibnamefont {Singha}}, \ and\
  \bibinfo {author} {\bibfnamefont {Emmanuel~N.}\ \bibnamefont {Saridakis}},\
  }\bibfield  {title} {\enquote {\bibinfo {title} {{Observational constraints
  on sign-changeable interaction models and alleviation of the $H_0$
  tension}},}\ }\href {\doibase 10.1103/PhysRevD.100.083539} {\bibfield
  {journal} {\bibinfo  {journal} {Phys. Rev. D}\ }\textbf {\bibinfo {volume}
  {100}},\ \bibinfo {pages} {083539} (\bibinfo {year} {2019}{\natexlab{a}})},\
  \Eprint {http://arxiv.org/abs/1903.10969} {arXiv:1903.10969 [astro-ph.CO]}
  \BibitemShut {NoStop}%
\bibitem [{\citenamefont {Pan}\ \emph {et~al.}(2019{\natexlab{b}})\citenamefont
  {Pan}, \citenamefont {Yang}, \citenamefont {Di~Valentino}, \citenamefont
  {Saridakis},\ and\ \citenamefont {Chakraborty}}]{Pan:2019gop}%
  \BibitemOpen
  \bibfield  {author} {\bibinfo {author} {\bibfnamefont {Supriya}\ \bibnamefont
  {Pan}}, \bibinfo {author} {\bibfnamefont {Weiqiang}\ \bibnamefont {Yang}},
  \bibinfo {author} {\bibfnamefont {Eleonora}\ \bibnamefont {Di~Valentino}},
  \bibinfo {author} {\bibfnamefont {Emmanuel~N.}\ \bibnamefont {Saridakis}}, \
  and\ \bibinfo {author} {\bibfnamefont {Subenoy}\ \bibnamefont
  {Chakraborty}},\ }\bibfield  {title} {\enquote {\bibinfo {title}
  {{Interacting scenarios with dynamical dark energy: Observational constraints
  and alleviation of the $H_0$ tension}},}\ }\href {\doibase
  10.1103/PhysRevD.100.103520} {\bibfield  {journal} {\bibinfo  {journal}
  {Phys. Rev. D}\ }\textbf {\bibinfo {volume} {100}},\ \bibinfo {pages}
  {103520} (\bibinfo {year} {2019}{\natexlab{b}})},\ \Eprint
  {http://arxiv.org/abs/1907.07540} {arXiv:1907.07540 [astro-ph.CO]}
  \BibitemShut {NoStop}%
\bibitem [{\citenamefont {Cheng}\ \emph {et~al.}(2020)\citenamefont {Cheng},
  \citenamefont {Ma}, \citenamefont {Wu}, \citenamefont {Zhang},\ and\
  \citenamefont {Chen}}]{Cheng:2019bkh}%
  \BibitemOpen
  \bibfield  {author} {\bibinfo {author} {\bibfnamefont {Gong}\ \bibnamefont
  {Cheng}}, \bibinfo {author} {\bibfnamefont {Yin-Zhe}\ \bibnamefont {Ma}},
  \bibinfo {author} {\bibfnamefont {Fengquan}\ \bibnamefont {Wu}}, \bibinfo
  {author} {\bibfnamefont {Jiajun}\ \bibnamefont {Zhang}}, \ and\ \bibinfo
  {author} {\bibfnamefont {Xuelei}\ \bibnamefont {Chen}},\ }\bibfield  {title}
  {\enquote {\bibinfo {title} {{Testing interacting dark matter and dark energy
  model with cosmological data}},}\ }\href {\doibase
  10.1103/PhysRevD.102.043517} {\bibfield  {journal} {\bibinfo  {journal}
  {Phys. Rev. D}\ }\textbf {\bibinfo {volume} {102}},\ \bibinfo {pages}
  {043517} (\bibinfo {year} {2020})},\ \Eprint
  {http://arxiv.org/abs/1911.04520} {arXiv:1911.04520 [astro-ph.CO]}
  \BibitemShut {NoStop}%
\bibitem [{\citenamefont {Pan}\ \emph {et~al.}(2020)\citenamefont {Pan},
  \citenamefont {Sharov},\ and\ \citenamefont {Yang}}]{Pan:2020zza}%
  \BibitemOpen
  \bibfield  {author} {\bibinfo {author} {\bibfnamefont {Supriya}\ \bibnamefont
  {Pan}}, \bibinfo {author} {\bibfnamefont {German~S.}\ \bibnamefont {Sharov}},
  \ and\ \bibinfo {author} {\bibfnamefont {Weiqiang}\ \bibnamefont {Yang}},\
  }\bibfield  {title} {\enquote {\bibinfo {title} {{Field theoretic
  interpretations of interacting dark energy scenarios and recent
  observations}},}\ }\href {\doibase 10.1103/PhysRevD.101.103533} {\bibfield
  {journal} {\bibinfo  {journal} {Phys. Rev. D}\ }\textbf {\bibinfo {volume}
  {101}},\ \bibinfo {pages} {103533} (\bibinfo {year} {2020})},\ \Eprint
  {http://arxiv.org/abs/2001.03120} {arXiv:2001.03120 [astro-ph.CO]}
  \BibitemShut {NoStop}%
\bibitem [{\citenamefont {Yang}\ \emph {et~al.}(2021)\citenamefont {Yang},
  \citenamefont {Pan}, \citenamefont {Arest\'e~Sal\'o},\ and\ \citenamefont
  {de~Haro}}]{Yang:2021oxc}%
  \BibitemOpen
  \bibfield  {author} {\bibinfo {author} {\bibfnamefont {Weiqiang}\
  \bibnamefont {Yang}}, \bibinfo {author} {\bibfnamefont {Supriya}\
  \bibnamefont {Pan}}, \bibinfo {author} {\bibfnamefont {Llibert}\ \bibnamefont
  {Arest\'e~Sal\'o}}, \ and\ \bibinfo {author} {\bibfnamefont {Jaume}\
  \bibnamefont {de~Haro}},\ }\bibfield  {title} {\enquote {\bibinfo {title}
  {{Theoretical and observational bounds on some interacting vacuum energy
  scenarios}},}\ }\href {\doibase 10.1103/PhysRevD.103.083520} {\bibfield
  {journal} {\bibinfo  {journal} {Phys. Rev. D}\ }\textbf {\bibinfo {volume}
  {103}},\ \bibinfo {pages} {083520} (\bibinfo {year} {2021})},\ \Eprint
  {http://arxiv.org/abs/2104.04505} {arXiv:2104.04505 [astro-ph.CO]}
  \BibitemShut {NoStop}%
\bibitem [{\citenamefont {Dutta}\ \emph {et~al.}(2018)\citenamefont {Dutta},
  \citenamefont {Khyllep},\ and\ \citenamefont {Tamanini}}]{Dutta:2017wfd}%
  \BibitemOpen
  \bibfield  {author} {\bibinfo {author} {\bibfnamefont {Jibitesh}\
  \bibnamefont {Dutta}}, \bibinfo {author} {\bibfnamefont {Wompherdeiki}\
  \bibnamefont {Khyllep}}, \ and\ \bibinfo {author} {\bibfnamefont {Nicola}\
  \bibnamefont {Tamanini}},\ }\bibfield  {title} {\enquote {\bibinfo {title}
  {{Dark energy with a gradient coupling to the dark matter fluid: cosmological
  dynamics and structure formation}},}\ }\href {\doibase
  10.1088/1475-7516/2018/01/038} {\bibfield  {journal} {\bibinfo  {journal}
  {JCAP}\ }\textbf {\bibinfo {volume} {01}},\ \bibinfo {pages} {038} (\bibinfo
  {year} {2018})},\ \Eprint {http://arxiv.org/abs/1707.09246} {arXiv:1707.09246
  [gr-qc]} \BibitemShut {NoStop}%
\bibitem [{\citenamefont {Khyllep}\ and\ \citenamefont
  {Dutta}(2019)}]{Khyllep:2019odd}%
  \BibitemOpen
  \bibfield  {author} {\bibinfo {author} {\bibfnamefont {Wompherdeiki}\
  \bibnamefont {Khyllep}}\ and\ \bibinfo {author} {\bibfnamefont {Jibitesh}\
  \bibnamefont {Dutta}},\ }\bibfield  {title} {\enquote {\bibinfo {title}
  {{Linear growth index of matter perturbations in Rastall gravity}},}\ }\href
  {\doibase 10.1016/j.physletb.2019.134796} {\bibfield  {journal} {\bibinfo
  {journal} {Phys. Lett. B}\ }\textbf {\bibinfo {volume} {797}},\ \bibinfo
  {pages} {134796} (\bibinfo {year} {2019})},\ \Eprint
  {http://arxiv.org/abs/1907.09221} {arXiv:1907.09221 [gr-qc]} \BibitemShut
  {NoStop}%
\bibitem [{\citenamefont {Khyllep}\ \emph {et~al.}(2021)\citenamefont
  {Khyllep}, \citenamefont {Paliathanasis},\ and\ \citenamefont
  {Dutta}}]{Khyllep:2021pcu}%
  \BibitemOpen
  \bibfield  {author} {\bibinfo {author} {\bibfnamefont {Wompherdeiki}\
  \bibnamefont {Khyllep}}, \bibinfo {author} {\bibfnamefont {Andronikos}\
  \bibnamefont {Paliathanasis}}, \ and\ \bibinfo {author} {\bibfnamefont
  {Jibitesh}\ \bibnamefont {Dutta}},\ }\bibfield  {title} {\enquote {\bibinfo
  {title} {{Cosmological solutions and growth index of matter perturbations in
  $f(Q)$ gravity}},}\ }\href {\doibase 10.1103/PhysRevD.103.103521} {\bibfield
  {journal} {\bibinfo  {journal} {Phys. Rev. D}\ }\textbf {\bibinfo {volume}
  {103}},\ \bibinfo {pages} {103521} (\bibinfo {year} {2021})},\ \Eprint
  {http://arxiv.org/abs/2103.08372} {arXiv:2103.08372 [gr-qc]} \BibitemShut
  {NoStop}%
\bibitem [{\citenamefont {Paliathanasis}\ \emph {et~al.}(2021)\citenamefont
  {Paliathanasis}, \citenamefont {Leon}, \citenamefont {Khyllep}, \citenamefont
  {Dutta},\ and\ \citenamefont {Pan}}]{Paliathanasis:2021egx}%
  \BibitemOpen
  \bibfield  {author} {\bibinfo {author} {\bibfnamefont {Andronikos}\
  \bibnamefont {Paliathanasis}}, \bibinfo {author} {\bibfnamefont {Genly}\
  \bibnamefont {Leon}}, \bibinfo {author} {\bibfnamefont {Wompherdeiki}\
  \bibnamefont {Khyllep}}, \bibinfo {author} {\bibfnamefont {Jibitesh}\
  \bibnamefont {Dutta}}, \ and\ \bibinfo {author} {\bibfnamefont {Supriya}\
  \bibnamefont {Pan}},\ }\bibfield  {title} {\enquote {\bibinfo {title}
  {{Interacting quintessence in light of generalized uncertainty principle:
  cosmological perturbations and dynamics}},}\ }\href {\doibase
  10.1140/epjc/s10052-021-09362-8} {\bibfield  {journal} {\bibinfo  {journal}
  {Eur. Phys. J. C}\ }\textbf {\bibinfo {volume} {81}},\ \bibinfo {pages} {607}
  (\bibinfo {year} {2021})},\ \Eprint {http://arxiv.org/abs/2104.06097}
  {arXiv:2104.06097 [gr-qc]} \BibitemShut {NoStop}%
\bibitem [{\citenamefont {Caldera-Cabral}\ \emph {et~al.}(2009)\citenamefont
  {Caldera-Cabral}, \citenamefont {Maartens},\ and\ \citenamefont
  {Schaefer}}]{CalderaCabral:2009ja}%
  \BibitemOpen
  \bibfield  {author} {\bibinfo {author} {\bibfnamefont {Gabriela}\
  \bibnamefont {Caldera-Cabral}}, \bibinfo {author} {\bibfnamefont {Roy}\
  \bibnamefont {Maartens}}, \ and\ \bibinfo {author} {\bibfnamefont
  {Bjoern~Malte}\ \bibnamefont {Schaefer}},\ }\bibfield  {title} {\enquote
  {\bibinfo {title} {{The Growth of Structure in Interacting Dark Energy
  Models}},}\ }\href {\doibase 10.1088/1475-7516/2009/07/027} {\bibfield
  {journal} {\bibinfo  {journal} {JCAP}\ }\textbf {\bibinfo {volume} {07}},\
  \bibinfo {pages} {027} (\bibinfo {year} {2009})},\ \Eprint
  {http://arxiv.org/abs/0905.0492} {arXiv:0905.0492 [astro-ph.CO]} \BibitemShut
  {NoStop}%
\bibitem [{\citenamefont {Tsujikawa}\ \emph {et~al.}(2013)\citenamefont
  {Tsujikawa}, \citenamefont {De~Felice},\ and\ \citenamefont
  {Alcaniz}}]{Tsujikawa:2012hv}%
  \BibitemOpen
  \bibfield  {author} {\bibinfo {author} {\bibfnamefont {Shinji}\ \bibnamefont
  {Tsujikawa}}, \bibinfo {author} {\bibfnamefont {Antonio}\ \bibnamefont
  {De~Felice}}, \ and\ \bibinfo {author} {\bibfnamefont {Jailson}\ \bibnamefont
  {Alcaniz}},\ }\bibfield  {title} {\enquote {\bibinfo {title} {{Testing for
  dynamical dark energy models with redshift-space distortions}},}\ }\href
  {\doibase 10.1088/1475-7516/2013/01/030} {\bibfield  {journal} {\bibinfo
  {journal} {JCAP}\ }\textbf {\bibinfo {volume} {01}},\ \bibinfo {pages} {030}
  (\bibinfo {year} {2013})},\ \Eprint {http://arxiv.org/abs/1210.4239}
  {arXiv:1210.4239 [astro-ph.CO]} \BibitemShut {NoStop}%
\bibitem [{\citenamefont {Li}\ \emph {et~al.}(2014)\citenamefont {Li},
  \citenamefont {Zhang},\ and\ \citenamefont {Zhang}}]{PhysRevD.90.123007}%
  \BibitemOpen
  \bibfield  {author} {\bibinfo {author} {\bibfnamefont {Yun-He}\ \bibnamefont
  {Li}}, \bibinfo {author} {\bibfnamefont {Jing-Fei}\ \bibnamefont {Zhang}}, \
  and\ \bibinfo {author} {\bibfnamefont {Xin}\ \bibnamefont {Zhang}},\
  }\bibfield  {title} {\enquote {\bibinfo {title} {Exploring the full parameter
  space for an interacting dark energy model with recent observations including
  redshift-space distortions: Application of the parametrized post-friedmann
  approach},}\ }\href {\doibase 10.1103/PhysRevD.90.123007} {\bibfield
  {journal} {\bibinfo  {journal} {Phys. Rev. D}\ }\textbf {\bibinfo {volume}
  {90}},\ \bibinfo {pages} {123007} (\bibinfo {year} {2014})}\BibitemShut
  {NoStop}%
\bibitem [{\citenamefont {Wainwright}\ and\ \citenamefont
  {Ellis}(1997)}]{wainwrightellis1997}%
  \BibitemOpen
  \bibfield  {author} {\bibinfo {author} {\bibfnamefont {John}\ \bibnamefont
  {Wainwright}}\ and\ \bibinfo {author} {\bibfnamefont {George F.~R.}\
  \bibnamefont {Ellis}},\ }\href {\doibase 10.1017/CBO9780511524660} {\emph
  {\bibinfo {title} {Dynamical Systems in Cosmology}}}\ (\bibinfo  {publisher}
  {Cambridge University Press (Cambridge)},\ \bibinfo {year}
  {1997})\BibitemShut {NoStop}%
\bibitem [{\citenamefont {Coley}(2003)}]{Coley:2003mj}%
  \BibitemOpen
  \bibfield  {author} {\bibinfo {author} {\bibfnamefont {A.~A.}\ \bibnamefont
  {Coley}},\ }\href {\doibase 10.1007/978-94-017-0327-7} {\emph {\bibinfo
  {title} {{Dynamical systems and cosmology}}}}\ (\bibinfo  {publisher}
  {Kluwer},\ \bibinfo {address} {Dordrecht, Netherlands},\ \bibinfo {year}
  {2003})\BibitemShut {NoStop}%
\bibitem [{\citenamefont {Bahamonde}\ \emph {et~al.}(2018)\citenamefont
  {Bahamonde}, \citenamefont {B\"ohmer}, \citenamefont {Carloni}, \citenamefont
  {Copeland}, \citenamefont {Fang},\ and\ \citenamefont
  {Tamanini}}]{Bahamonde:2017ize}%
  \BibitemOpen
  \bibfield  {author} {\bibinfo {author} {\bibfnamefont {Sebastian}\
  \bibnamefont {Bahamonde}}, \bibinfo {author} {\bibfnamefont {Christian~G.}\
  \bibnamefont {B\"ohmer}}, \bibinfo {author} {\bibfnamefont {Sante}\
  \bibnamefont {Carloni}}, \bibinfo {author} {\bibfnamefont {Edmund~J.}\
  \bibnamefont {Copeland}}, \bibinfo {author} {\bibfnamefont {Wei}\
  \bibnamefont {Fang}}, \ and\ \bibinfo {author} {\bibfnamefont {Nicola}\
  \bibnamefont {Tamanini}},\ }\bibfield  {title} {\enquote {\bibinfo {title}
  {{Dynamical systems applied to cosmology: dark energy and modified
  gravity}},}\ }\href {\doibase 10.1016/j.physrep.2018.09.001} {\bibfield
  {journal} {\bibinfo  {journal} {Phys. Rept.}\ }\textbf {\bibinfo {volume}
  {775-777}},\ \bibinfo {pages} {1--122} (\bibinfo {year} {2018})},\ \Eprint
  {http://arxiv.org/abs/1712.03107} {arXiv:1712.03107 [gr-qc]} \BibitemShut
  {NoStop}%
\bibitem [{\citenamefont {Copeland}\ \emph {et~al.}(1998)\citenamefont
  {Copeland}, \citenamefont {Liddle},\ and\ \citenamefont
  {Wands}}]{Copeland:1997et}%
  \BibitemOpen
  \bibfield  {author} {\bibinfo {author} {\bibfnamefont {Edmund~J.}\
  \bibnamefont {Copeland}}, \bibinfo {author} {\bibfnamefont {Andrew~R}\
  \bibnamefont {Liddle}}, \ and\ \bibinfo {author} {\bibfnamefont {David}\
  \bibnamefont {Wands}},\ }\bibfield  {title} {\enquote {\bibinfo {title}
  {{Exponential potentials and cosmological scaling solutions}},}\ }\href
  {\doibase 10.1103/PhysRevD.57.4686} {\bibfield  {journal} {\bibinfo
  {journal} {Phys. Rev. D}\ }\textbf {\bibinfo {volume} {57}},\ \bibinfo
  {pages} {4686--4690} (\bibinfo {year} {1998})},\ \Eprint
  {http://arxiv.org/abs/gr-qc/9711068} {arXiv:gr-qc/9711068} \BibitemShut
  {NoStop}%
\bibitem [{\citenamefont {Gong}\ \emph {et~al.}(2006)\citenamefont {Gong},
  \citenamefont {Wang},\ and\ \citenamefont {Zhang}}]{Gong:2006sp}%
  \BibitemOpen
  \bibfield  {author} {\bibinfo {author} {\bibfnamefont {Yungui}\ \bibnamefont
  {Gong}}, \bibinfo {author} {\bibfnamefont {Anzhong}\ \bibnamefont {Wang}}, \
  and\ \bibinfo {author} {\bibfnamefont {Yuan-Zhong}\ \bibnamefont {Zhang}},\
  }\bibfield  {title} {\enquote {\bibinfo {title} {{Exact scaling solutions and
  fixed points for general scalar field}},}\ }\href {\doibase
  10.1016/j.physletb.2006.03.057} {\bibfield  {journal} {\bibinfo  {journal}
  {Phys. Lett. B}\ }\textbf {\bibinfo {volume} {636}},\ \bibinfo {pages}
  {286--292} (\bibinfo {year} {2006})},\ \Eprint
  {http://arxiv.org/abs/gr-qc/0603050} {arXiv:gr-qc/0603050} \BibitemShut
  {NoStop}%
\bibitem [{\citenamefont {Setare}\ and\ \citenamefont
  {Saridakis}(2009)}]{Setare:2008sf}%
  \BibitemOpen
  \bibfield  {author} {\bibinfo {author} {\bibfnamefont {M.~R.}\ \bibnamefont
  {Setare}}\ and\ \bibinfo {author} {\bibfnamefont {E.~N.}\ \bibnamefont
  {Saridakis}},\ }\bibfield  {title} {\enquote {\bibinfo {title} {{Quintom dark
  energy models with nearly flat potentials}},}\ }\href {\doibase
  10.1103/PhysRevD.79.043005} {\bibfield  {journal} {\bibinfo  {journal} {Phys.
  Rev. D}\ }\textbf {\bibinfo {volume} {79}},\ \bibinfo {pages} {043005}
  (\bibinfo {year} {2009})},\ \Eprint {http://arxiv.org/abs/0810.4775}
  {arXiv:0810.4775 [astro-ph]} \BibitemShut {NoStop}%
\bibitem [{\citenamefont {Matos}\ \emph {et~al.}(2009)\citenamefont {Matos},
  \citenamefont {Luevano}, \citenamefont {Quiros}, \citenamefont
  {Urena-Lopez},\ and\ \citenamefont {Vazquez}}]{Matos:2009hf}%
  \BibitemOpen
  \bibfield  {author} {\bibinfo {author} {\bibfnamefont {Tonatiuh}\
  \bibnamefont {Matos}}, \bibinfo {author} {\bibfnamefont {Jose-Ruben}\
  \bibnamefont {Luevano}}, \bibinfo {author} {\bibfnamefont {Israel}\
  \bibnamefont {Quiros}}, \bibinfo {author} {\bibfnamefont {L.~Arturo}\
  \bibnamefont {Urena-Lopez}}, \ and\ \bibinfo {author} {\bibfnamefont
  {Jose~Alberto}\ \bibnamefont {Vazquez}},\ }\bibfield  {title} {\enquote
  {\bibinfo {title} {{Dynamics of Scalar Field Dark Matter With a Cosh-like
  Potential}},}\ }\href {\doibase 10.1103/PhysRevD.80.123521} {\bibfield
  {journal} {\bibinfo  {journal} {Phys. Rev. D}\ }\textbf {\bibinfo {volume}
  {80}},\ \bibinfo {pages} {123521} (\bibinfo {year} {2009})},\ \Eprint
  {http://arxiv.org/abs/0906.0396} {arXiv:0906.0396 [astro-ph.CO]} \BibitemShut
  {NoStop}%
\bibitem [{\citenamefont {Copeland}\ \emph {et~al.}(2009)\citenamefont
  {Copeland}, \citenamefont {Mizuno},\ and\ \citenamefont
  {Shaeri}}]{Copeland:2009be}%
  \BibitemOpen
  \bibfield  {author} {\bibinfo {author} {\bibfnamefont {Edmund~J.}\
  \bibnamefont {Copeland}}, \bibinfo {author} {\bibfnamefont {Shuntaro}\
  \bibnamefont {Mizuno}}, \ and\ \bibinfo {author} {\bibfnamefont {Maryam}\
  \bibnamefont {Shaeri}},\ }\bibfield  {title} {\enquote {\bibinfo {title}
  {{Dynamics of a scalar field in Robertson-Walker spacetimes}},}\ }\href
  {\doibase 10.1103/PhysRevD.79.103515} {\bibfield  {journal} {\bibinfo
  {journal} {Phys. Rev. D}\ }\textbf {\bibinfo {volume} {79}},\ \bibinfo
  {pages} {103515} (\bibinfo {year} {2009})},\ \Eprint
  {http://arxiv.org/abs/0904.0877} {arXiv:0904.0877 [astro-ph.CO]} \BibitemShut
  {NoStop}%
\bibitem [{\citenamefont {Leyva}\ \emph {et~al.}(2009)\citenamefont {Leyva},
  \citenamefont {Gonzalez}, \citenamefont {Gonzalez}, \citenamefont {Matos},\
  and\ \citenamefont {Quiros}}]{Leyva:2009zz}%
  \BibitemOpen
  \bibfield  {author} {\bibinfo {author} {\bibfnamefont {Yoelsy}\ \bibnamefont
  {Leyva}}, \bibinfo {author} {\bibfnamefont {Dania}\ \bibnamefont {Gonzalez}},
  \bibinfo {author} {\bibfnamefont {Tame}\ \bibnamefont {Gonzalez}}, \bibinfo
  {author} {\bibfnamefont {Tonatiuh}\ \bibnamefont {Matos}}, \ and\ \bibinfo
  {author} {\bibfnamefont {Israel}\ \bibnamefont {Quiros}},\ }\bibfield
  {title} {\enquote {\bibinfo {title} {{Dynamics of a self-interacting scalar
  field trapped in the braneworld for a wide variety of self-interaction
  potentials}},}\ }\href {\doibase 10.1103/PhysRevD.80.044026} {\bibfield
  {journal} {\bibinfo  {journal} {Phys. Rev. D}\ }\textbf {\bibinfo {volume}
  {80}},\ \bibinfo {pages} {044026} (\bibinfo {year} {2009})},\ \Eprint
  {http://arxiv.org/abs/0909.0281} {arXiv:0909.0281 [gr-qc]} \BibitemShut
  {NoStop}%
\bibitem [{\citenamefont {Leon}\ and\ \citenamefont
  {Saridakis}(2011)}]{Leon:2010pu}%
  \BibitemOpen
  \bibfield  {author} {\bibinfo {author} {\bibfnamefont {Genly}\ \bibnamefont
  {Leon}}\ and\ \bibinfo {author} {\bibfnamefont {Emmanuel~N.}\ \bibnamefont
  {Saridakis}},\ }\bibfield  {title} {\enquote {\bibinfo {title} {{Dynamics of
  the anisotropic Kantowsky-Sachs geometries in $R^n$ gravity}},}\ }\href
  {\doibase 10.1088/0264-9381/28/6/065008} {\bibfield  {journal} {\bibinfo
  {journal} {Class. Quant. Grav.}\ }\textbf {\bibinfo {volume} {28}},\ \bibinfo
  {pages} {065008} (\bibinfo {year} {2011})},\ \Eprint
  {http://arxiv.org/abs/1007.3956} {arXiv:1007.3956 [gr-qc]} \BibitemShut
  {NoStop}%
\bibitem [{\citenamefont {Urena-Lopez}(2012)}]{Urena-Lopez:2011gxx}%
  \BibitemOpen
  \bibfield  {author} {\bibinfo {author} {\bibfnamefont {L.~Arturo}\
  \bibnamefont {Urena-Lopez}},\ }\bibfield  {title} {\enquote {\bibinfo {title}
  {{Unified description of the dynamics of quintessential scalar fields}},}\
  }\href {\doibase 10.1088/1475-7516/2012/03/035} {\bibfield  {journal}
  {\bibinfo  {journal} {JCAP}\ }\textbf {\bibinfo {volume} {03}},\ \bibinfo
  {pages} {035} (\bibinfo {year} {2012})},\ \Eprint
  {http://arxiv.org/abs/1108.4712} {arXiv:1108.4712 [astro-ph.CO]} \BibitemShut
  {NoStop}%
\bibitem [{\citenamefont {Leon}\ \emph {et~al.}(2013)\citenamefont {Leon},
  \citenamefont {Saavedra},\ and\ \citenamefont {Saridakis}}]{Leon:2013qh}%
  \BibitemOpen
  \bibfield  {author} {\bibinfo {author} {\bibfnamefont {Genly}\ \bibnamefont
  {Leon}}, \bibinfo {author} {\bibfnamefont {Joel}\ \bibnamefont {Saavedra}}, \
  and\ \bibinfo {author} {\bibfnamefont {Emmanuel~N.}\ \bibnamefont
  {Saridakis}},\ }\bibfield  {title} {\enquote {\bibinfo {title} {{Cosmological
  behavior in extended nonlinear massive gravity}},}\ }\href {\doibase
  10.1088/0264-9381/30/13/135001} {\bibfield  {journal} {\bibinfo  {journal}
  {Class. Quant. Grav.}\ }\textbf {\bibinfo {volume} {30}},\ \bibinfo {pages}
  {135001} (\bibinfo {year} {2013})},\ \Eprint {http://arxiv.org/abs/1301.7419}
  {arXiv:1301.7419 [astro-ph.CO]} \BibitemShut {NoStop}%
\bibitem [{\citenamefont {Fadragas}\ \emph {et~al.}(2014)\citenamefont
  {Fadragas}, \citenamefont {Leon},\ and\ \citenamefont
  {Saridakis}}]{Fadragas:2013ina}%
  \BibitemOpen
  \bibfield  {author} {\bibinfo {author} {\bibfnamefont {Carlos~R.}\
  \bibnamefont {Fadragas}}, \bibinfo {author} {\bibfnamefont {Genly}\
  \bibnamefont {Leon}}, \ and\ \bibinfo {author} {\bibfnamefont {Emmanuel~N.}\
  \bibnamefont {Saridakis}},\ }\bibfield  {title} {\enquote {\bibinfo {title}
  {{Dynamical analysis of anisotropic scalar-field cosmologies for a wide range
  of potentials}},}\ }\href {\doibase 10.1088/0264-9381/31/7/075018} {\bibfield
   {journal} {\bibinfo  {journal} {Class. Quant. Grav.}\ }\textbf {\bibinfo
  {volume} {31}},\ \bibinfo {pages} {075018} (\bibinfo {year} {2014})},\
  \Eprint {http://arxiv.org/abs/1308.1658} {arXiv:1308.1658 [gr-qc]}
  \BibitemShut {NoStop}%
\bibitem [{\citenamefont {Khyllep}\ and\ \citenamefont
  {Dutta}(2021)}]{Khyllep:2021yyp}%
  \BibitemOpen
  \bibfield  {author} {\bibinfo {author} {\bibfnamefont {Wompherdeiki}\
  \bibnamefont {Khyllep}}\ and\ \bibinfo {author} {\bibfnamefont {Jibitesh}\
  \bibnamefont {Dutta}},\ }\bibfield  {title} {\enquote {\bibinfo {title}
  {{Cosmological dynamics and bifurcation analysis of the general non-minimal
  coupled scalar field models}},}\ }\href {\doibase
  10.1140/epjc/s10052-021-09559-x} {\bibfield  {journal} {\bibinfo  {journal}
  {Eur. Phys. J. C}\ }\textbf {\bibinfo {volume} {81}},\ \bibinfo {pages} {774}
  (\bibinfo {year} {2021})},\ \Eprint {http://arxiv.org/abs/2102.04744}
  {arXiv:2102.04744 [gr-qc]} \BibitemShut {NoStop}%
\bibitem [{\citenamefont {Zonunmawia}\ \emph {et~al.}(2018)\citenamefont
  {Zonunmawia}, \citenamefont {Khyllep}, \citenamefont {Dutta},\ and\
  \citenamefont {J\"arv}}]{Zonunmawia:2018xvf}%
  \BibitemOpen
  \bibfield  {author} {\bibinfo {author} {\bibfnamefont {Hmar}\ \bibnamefont
  {Zonunmawia}}, \bibinfo {author} {\bibfnamefont {Wompherdeiki}\ \bibnamefont
  {Khyllep}}, \bibinfo {author} {\bibfnamefont {Jibitesh}\ \bibnamefont
  {Dutta}}, \ and\ \bibinfo {author} {\bibfnamefont {Laur}\ \bibnamefont
  {J\"arv}},\ }\bibfield  {title} {\enquote {\bibinfo {title} {{Cosmological
  dynamics of brane gravity: A global dynamical system perspective}},}\ }\href
  {\doibase 10.1103/PhysRevD.98.083532} {\bibfield  {journal} {\bibinfo
  {journal} {Phys. Rev. D}\ }\textbf {\bibinfo {volume} {98}},\ \bibinfo
  {pages} {083532} (\bibinfo {year} {2018})},\ \Eprint
  {http://arxiv.org/abs/1810.03816} {arXiv:1810.03816 [gr-qc]} \BibitemShut
  {NoStop}%
\bibitem [{\citenamefont {Dutta}\ \emph {et~al.}(2017)\citenamefont {Dutta},
  \citenamefont {Khyllep},\ and\ \citenamefont {Tamanini}}]{Dutta:2017kch}%
  \BibitemOpen
  \bibfield  {author} {\bibinfo {author} {\bibfnamefont {Jibitesh}\
  \bibnamefont {Dutta}}, \bibinfo {author} {\bibfnamefont {Wompherdeiki}\
  \bibnamefont {Khyllep}}, \ and\ \bibinfo {author} {\bibfnamefont {Nicola}\
  \bibnamefont {Tamanini}},\ }\bibfield  {title} {\enquote {\bibinfo {title}
  {{Scalar-Fluid interacting dark energy: cosmological dynamics beyond the
  exponential potential}},}\ }\href {\doibase 10.1103/PhysRevD.95.023515}
  {\bibfield  {journal} {\bibinfo  {journal} {Phys. Rev. D}\ }\textbf {\bibinfo
  {volume} {95}},\ \bibinfo {pages} {023515} (\bibinfo {year} {2017})},\
  \Eprint {http://arxiv.org/abs/1701.00744} {arXiv:1701.00744 [gr-qc]}
  \BibitemShut {NoStop}%
\bibitem [{\citenamefont {Dutta}\ \emph {et~al.}(2016)\citenamefont {Dutta},
  \citenamefont {Khyllep},\ and\ \citenamefont {Tamanini}}]{Dutta:2016bbs}%
  \BibitemOpen
  \bibfield  {author} {\bibinfo {author} {\bibfnamefont {Jibitesh}\
  \bibnamefont {Dutta}}, \bibinfo {author} {\bibfnamefont {Wompherdeiki}\
  \bibnamefont {Khyllep}}, \ and\ \bibinfo {author} {\bibfnamefont {Nicola}\
  \bibnamefont {Tamanini}},\ }\bibfield  {title} {\enquote {\bibinfo {title}
  {{Cosmological dynamics of scalar fields with kinetic corrections: Beyond the
  exponential potential}},}\ }\href {\doibase 10.1103/PhysRevD.93.063004}
  {\bibfield  {journal} {\bibinfo  {journal} {Phys. Rev. D}\ }\textbf {\bibinfo
  {volume} {93}},\ \bibinfo {pages} {063004} (\bibinfo {year} {2016})},\
  \Eprint {http://arxiv.org/abs/1602.06113} {arXiv:1602.06113 [gr-qc]}
  \BibitemShut {NoStop}%
\bibitem [{\citenamefont {Billyard}\ and\ \citenamefont
  {Coley}(2000)}]{Billyard:2000bh}%
  \BibitemOpen
  \bibfield  {author} {\bibinfo {author} {\bibfnamefont {Andrew~P.}\
  \bibnamefont {Billyard}}\ and\ \bibinfo {author} {\bibfnamefont {Alan~A.}\
  \bibnamefont {Coley}},\ }\bibfield  {title} {\enquote {\bibinfo {title}
  {{Interactions in scalar field cosmology}},}\ }\href {\doibase
  10.1103/PhysRevD.61.083503} {\bibfield  {journal} {\bibinfo  {journal} {Phys.
  Rev. D}\ }\textbf {\bibinfo {volume} {61}},\ \bibinfo {pages} {083503}
  (\bibinfo {year} {2000})},\ \Eprint {http://arxiv.org/abs/astro-ph/9908224}
  {arXiv:astro-ph/9908224} \BibitemShut {NoStop}%
\bibitem [{\citenamefont {Chen}\ \emph {et~al.}(2009)\citenamefont {Chen},
  \citenamefont {Gong},\ and\ \citenamefont {Saridakis}}]{Chen:2008ft}%
  \BibitemOpen
  \bibfield  {author} {\bibinfo {author} {\bibfnamefont {Xi-ming}\ \bibnamefont
  {Chen}}, \bibinfo {author} {\bibfnamefont {Yun-gui}\ \bibnamefont {Gong}}, \
  and\ \bibinfo {author} {\bibfnamefont {Emmanuel~N.}\ \bibnamefont
  {Saridakis}},\ }\bibfield  {title} {\enquote {\bibinfo {title} {{Phase-space
  analysis of interacting phantom cosmology}},}\ }\href {\doibase
  10.1088/1475-7516/2009/04/001} {\bibfield  {journal} {\bibinfo  {journal}
  {JCAP}\ }\textbf {\bibinfo {volume} {04}},\ \bibinfo {pages} {001} (\bibinfo
  {year} {2009})},\ \Eprint {http://arxiv.org/abs/0812.1117} {arXiv:0812.1117
  [gr-qc]} \BibitemShut {NoStop}%
\bibitem [{\citenamefont {Boehmer}\ \emph {et~al.}(2008)\citenamefont
  {Boehmer}, \citenamefont {Caldera-Cabral}, \citenamefont {Lazkoz},\ and\
  \citenamefont {Maartens}}]{Boehmer:2008av}%
  \BibitemOpen
  \bibfield  {author} {\bibinfo {author} {\bibfnamefont {Christian~G.}\
  \bibnamefont {Boehmer}}, \bibinfo {author} {\bibfnamefont {Gabriela}\
  \bibnamefont {Caldera-Cabral}}, \bibinfo {author} {\bibfnamefont {Ruth}\
  \bibnamefont {Lazkoz}}, \ and\ \bibinfo {author} {\bibfnamefont {Roy}\
  \bibnamefont {Maartens}},\ }\bibfield  {title} {\enquote {\bibinfo {title}
  {{Dynamics of dark energy with a coupling to dark matter}},}\ }\href
  {\doibase 10.1103/PhysRevD.78.023505} {\bibfield  {journal} {\bibinfo
  {journal} {Phys. Rev. D}\ }\textbf {\bibinfo {volume} {78}},\ \bibinfo
  {pages} {023505} (\bibinfo {year} {2008})},\ \Eprint
  {http://arxiv.org/abs/0801.1565} {arXiv:0801.1565 [gr-qc]} \BibitemShut
  {NoStop}%
\bibitem [{\citenamefont
  {Woszczyna}(1992{\natexlab{a}})}]{1992MNRAS.255..701W}%
  \BibitemOpen
  \bibfield  {author} {\bibinfo {author} {\bibfnamefont {Andrzej}\ \bibnamefont
  {Woszczyna}},\ }\bibfield  {title} {\enquote {\bibinfo {title} {{A dynamical
  systems approach to cosmological structure formation - Newtonian
  universe}},}\ }\href {\doibase 10.1093/mnras/255.4.701} {\bibfield  {journal}
  {\bibinfo  {journal} {MNRAS}\ }\textbf {\bibinfo {volume} {255}},\ \bibinfo
  {pages} {701--706} (\bibinfo {year} {1992}{\natexlab{a}})}\BibitemShut
  {NoStop}%
\bibitem [{\citenamefont {Woszczyna}(1992{\natexlab{b}})}]{Woszczyna:1988sc}%
  \BibitemOpen
  \bibfield  {author} {\bibinfo {author} {\bibfnamefont {Andrzej}\ \bibnamefont
  {Woszczyna}},\ }\bibfield  {title} {\enquote {\bibinfo {title} {{Gauge
  invariant cosmic structures: A Dynamic systems approach}},}\ }\href {\doibase
  10.1103/PhysRevD.45.1982} {\bibfield  {journal} {\bibinfo  {journal} {Phys.
  Rev. D}\ }\textbf {\bibinfo {volume} {45}},\ \bibinfo {pages} {1982--1988}
  (\bibinfo {year} {1992}{\natexlab{b}})}\BibitemShut {NoStop}%
\bibitem [{\citenamefont {Bruni}(1993)}]{PhysRevD.47.738}%
  \BibitemOpen
  \bibfield  {author} {\bibinfo {author} {\bibfnamefont {Marco}\ \bibnamefont
  {Bruni}},\ }\bibfield  {title} {\enquote {\bibinfo {title} {Stability of open
  universes},}\ }\href {\doibase 10.1103/PhysRevD.47.738} {\bibfield  {journal}
  {\bibinfo  {journal} {Phys. Rev. D}\ }\textbf {\bibinfo {volume} {47}},\
  \bibinfo {pages} {738--742} (\bibinfo {year} {1993})}\BibitemShut {NoStop}%
\bibitem [{\citenamefont {Bruni}\ and\ \citenamefont
  {Piotrkowska}(1994)}]{Bruni:1993da}%
  \BibitemOpen
  \bibfield  {author} {\bibinfo {author} {\bibfnamefont {Marco}\ \bibnamefont
  {Bruni}}\ and\ \bibinfo {author} {\bibfnamefont {Kamilla}\ \bibnamefont
  {Piotrkowska}},\ }\bibfield  {title} {\enquote {\bibinfo {title} {{Dust -
  radiation universes: Stability analysis}},}\ }\href {\doibase
  10.1093/mnras/270.3.630} {\bibfield  {journal} {\bibinfo  {journal} {Mon.
  Not. Roy. Astron. Soc.}\ }\textbf {\bibinfo {volume} {270}},\ \bibinfo
  {pages} {630--640} (\bibinfo {year} {1994})},\ \Eprint
  {http://arxiv.org/abs/astro-ph/9308004} {arXiv:astro-ph/9308004} \BibitemShut
  {NoStop}%
\bibitem [{\citenamefont {Dunsby}(1993)}]{PhysRevD.48.3562}%
  \BibitemOpen
  \bibfield  {author} {\bibinfo {author} {\bibfnamefont {Peter K.~S.}\
  \bibnamefont {Dunsby}},\ }\bibfield  {title} {\enquote {\bibinfo {title}
  {Covariant perturbations of anisotropic cosmological models},}\ }\href
  {\doibase 10.1103/PhysRevD.48.3562} {\bibfield  {journal} {\bibinfo
  {journal} {Phys. Rev. D}\ }\textbf {\bibinfo {volume} {48}},\ \bibinfo
  {pages} {3562--3576} (\bibinfo {year} {1993})}\BibitemShut {NoStop}%
\bibitem [{\citenamefont {Hobbs}\ and\ \citenamefont
  {Dunsby}(2000)}]{PhysRevD.62.124007}%
  \BibitemOpen
  \bibfield  {author} {\bibinfo {author} {\bibfnamefont {Stacey}\ \bibnamefont
  {Hobbs}}\ and\ \bibinfo {author} {\bibfnamefont {Peter K.~S.}\ \bibnamefont
  {Dunsby}},\ }\bibfield  {title} {\enquote {\bibinfo {title} {Dynamical
  systems approach to magnetized cosmological perturbations},}\ }\href
  {\doibase 10.1103/PhysRevD.62.124007} {\bibfield  {journal} {\bibinfo
  {journal} {Phys. Rev. D}\ }\textbf {\bibinfo {volume} {62}},\ \bibinfo
  {pages} {124007} (\bibinfo {year} {2000})}\BibitemShut {NoStop}%
\bibitem [{\citenamefont {Alho}\ and\ \citenamefont
  {Mena}(2014)}]{PhysRevD.90.043501}%
  \BibitemOpen
  \bibfield  {author} {\bibinfo {author} {\bibfnamefont {Artur}\ \bibnamefont
  {Alho}}\ and\ \bibinfo {author} {\bibfnamefont {Filipe~C.}\ \bibnamefont
  {Mena}},\ }\bibfield  {title} {\enquote {\bibinfo {title} {Covariant and
  gauge-invariant linear scalar perturbations in multiple scalar field
  cosmologies},}\ }\href {\doibase 10.1103/PhysRevD.90.043501} {\bibfield
  {journal} {\bibinfo  {journal} {Phys. Rev. D}\ }\textbf {\bibinfo {volume}
  {90}},\ \bibinfo {pages} {043501} (\bibinfo {year} {2014})}\BibitemShut
  {NoStop}%
\bibitem [{\citenamefont {Basilakos}\ \emph {et~al.}(2019)\citenamefont
  {Basilakos}, \citenamefont {Leon}, \citenamefont {Papagiannopoulos},\ and\
  \citenamefont {Saridakis}}]{Basilakos:2019dof}%
  \BibitemOpen
  \bibfield  {author} {\bibinfo {author} {\bibfnamefont {Spyros}\ \bibnamefont
  {Basilakos}}, \bibinfo {author} {\bibfnamefont {Genly}\ \bibnamefont {Leon}},
  \bibinfo {author} {\bibfnamefont {G.}~\bibnamefont {Papagiannopoulos}}, \
  and\ \bibinfo {author} {\bibfnamefont {Emmanuel~N.}\ \bibnamefont
  {Saridakis}},\ }\bibfield  {title} {\enquote {\bibinfo {title} {{Dynamical
  system analysis at background and perturbation levels: Quintessence in severe
  disadvantage comparing to $\Lambda$CDM}},}\ }\href {\doibase
  10.1103/PhysRevD.100.043524} {\bibfield  {journal} {\bibinfo  {journal}
  {Phys. Rev. D}\ }\textbf {\bibinfo {volume} {100}},\ \bibinfo {pages}
  {043524} (\bibinfo {year} {2019})},\ \Eprint
  {http://arxiv.org/abs/1904.01563} {arXiv:1904.01563 [gr-qc]} \BibitemShut
  {NoStop}%
\bibitem [{\citenamefont {Alho}\ \emph {et~al.}(2019)\citenamefont {Alho},
  \citenamefont {Uggla},\ and\ \citenamefont {Wainwright}}]{Alho:2019jho}%
  \BibitemOpen
  \bibfield  {author} {\bibinfo {author} {\bibfnamefont {Artur}\ \bibnamefont
  {Alho}}, \bibinfo {author} {\bibfnamefont {Claes}\ \bibnamefont {Uggla}}, \
  and\ \bibinfo {author} {\bibfnamefont {John}\ \bibnamefont {Wainwright}},\
  }\bibfield  {title} {\enquote {\bibinfo {title} {{Perturbations of the
  Lambda-CDM model in a dynamical systems perspective}},}\ }\href {\doibase
  10.1088/1475-7516/2019/09/045} {\bibfield  {journal} {\bibinfo  {journal}
  {JCAP}\ }\textbf {\bibinfo {volume} {09}},\ \bibinfo {pages} {045} (\bibinfo
  {year} {2019})},\ \Eprint {http://arxiv.org/abs/1904.02463} {arXiv:1904.02463
  [gr-qc]} \BibitemShut {NoStop}%
\bibitem [{\citenamefont {Landim}(2019)}]{Landim:2019lvl}%
  \BibitemOpen
  \bibfield  {author} {\bibinfo {author} {\bibfnamefont {Ricardo~G.}\
  \bibnamefont {Landim}},\ }\bibfield  {title} {\enquote {\bibinfo {title}
  {{Cosmological perturbations and dynamical analysis for interacting
  quintessence}},}\ }\href {\doibase 10.1140/epjc/s10052-019-7418-8} {\bibfield
   {journal} {\bibinfo  {journal} {Eur. Phys. J. C}\ }\textbf {\bibinfo
  {volume} {79}},\ \bibinfo {pages} {889} (\bibinfo {year} {2019})},\ \Eprint
  {http://arxiv.org/abs/1908.03657} {arXiv:1908.03657 [gr-qc]} \BibitemShut
  {NoStop}%
\bibitem [{\citenamefont {Ma}\ and\ \citenamefont
  {Bertschinger}(1995)}]{Ma:1995ey}%
  \BibitemOpen
  \bibfield  {author} {\bibinfo {author} {\bibfnamefont {Chung-Pei}\
  \bibnamefont {Ma}}\ and\ \bibinfo {author} {\bibfnamefont {Edmund}\
  \bibnamefont {Bertschinger}},\ }\bibfield  {title} {\enquote {\bibinfo
  {title} {{Cosmological perturbation theory in the synchronous and conformal
  Newtonian gauges}},}\ }\href {\doibase 10.1086/176550} {\bibfield  {journal}
  {\bibinfo  {journal} {Astrophys. J.}\ }\textbf {\bibinfo {volume} {455}},\
  \bibinfo {pages} {7--25} (\bibinfo {year} {1995})},\ \Eprint
  {http://arxiv.org/abs/astro-ph/9506072} {arXiv:astro-ph/9506072} \BibitemShut
  {NoStop}%
\bibitem [{\citenamefont {Valiviita}\ \emph {et~al.}(2008)\citenamefont
  {Valiviita}, \citenamefont {Majerotto},\ and\ \citenamefont
  {Maartens}}]{Valiviita:2008iv}%
  \BibitemOpen
  \bibfield  {author} {\bibinfo {author} {\bibfnamefont {Jussi}\ \bibnamefont
  {Valiviita}}, \bibinfo {author} {\bibfnamefont {Elisabetta}\ \bibnamefont
  {Majerotto}}, \ and\ \bibinfo {author} {\bibfnamefont {Roy}\ \bibnamefont
  {Maartens}},\ }\bibfield  {title} {\enquote {\bibinfo {title} {{Instability
  in interacting dark energy and dark matter fluids}},}\ }\href {\doibase
  10.1088/1475-7516/2008/07/020} {\bibfield  {journal} {\bibinfo  {journal}
  {JCAP}\ }\textbf {\bibinfo {volume} {07}},\ \bibinfo {pages} {020} (\bibinfo
  {year} {2008})},\ \Eprint {http://arxiv.org/abs/0804.0232} {arXiv:0804.0232
  [astro-ph]} \BibitemShut {NoStop}%
\bibitem [{\citenamefont {Saridakis}(2021)}]{Saridakis:2021qxb}%
  \BibitemOpen
  \bibfield  {author} {\bibinfo {author} {\bibfnamefont {Emmanuel~N.}\
  \bibnamefont {Saridakis}},\ }\bibfield  {title} {\enquote {\bibinfo {title}
  {{Do we need soft cosmology?}}}\ }\href {\doibase
  10.1016/j.physletb.2021.136649} {\bibfield  {journal} {\bibinfo  {journal}
  {Phys. Lett. B}\ }\textbf {\bibinfo {volume} {822}},\ \bibinfo {pages}
  {136649} (\bibinfo {year} {2021})},\ \Eprint
  {http://arxiv.org/abs/2105.08646} {arXiv:2105.08646 [astro-ph.CO]}
  \BibitemShut {NoStop}%
\bibitem [{\citenamefont {Saridakis}\ \emph
  {et~al.}(2021{\natexlab{b}})\citenamefont {Saridakis}, \citenamefont {Yang},
  \citenamefont {Pan}, \citenamefont {Anagnostopoulos},\ and\ \citenamefont
  {Basilakos}}]{Saridakis:2021xqy}%
  \BibitemOpen
  \bibfield  {author} {\bibinfo {author} {\bibfnamefont {Emmanuel~N.}\
  \bibnamefont {Saridakis}}, \bibinfo {author} {\bibfnamefont {Weiqiang}\
  \bibnamefont {Yang}}, \bibinfo {author} {\bibfnamefont {Supriya}\
  \bibnamefont {Pan}}, \bibinfo {author} {\bibfnamefont {Fotios~K.}\
  \bibnamefont {Anagnostopoulos}}, \ and\ \bibinfo {author} {\bibfnamefont
  {Spyros}\ \bibnamefont {Basilakos}},\ }\bibfield  {title} {\enquote {\bibinfo
  {title} {{Observational constraints on soft dark energy and soft dark matter:
  challenging $\Lambda$CDM}},}\ }\href@noop {} {\  (\bibinfo {year}
  {2021}{\natexlab{b}})},\ \Eprint {http://arxiv.org/abs/2112.08330}
  {arXiv:2112.08330 [astro-ph.CO]} \BibitemShut {NoStop}%
\bibitem [{\citenamefont {He}\ \emph {et~al.}(2009)\citenamefont {He},
  \citenamefont {Wang},\ and\ \citenamefont {Abdalla}}]{He:2008si}%
  \BibitemOpen
  \bibfield  {author} {\bibinfo {author} {\bibfnamefont {Jian-Hua}\
  \bibnamefont {He}}, \bibinfo {author} {\bibfnamefont {Bin}\ \bibnamefont
  {Wang}}, \ and\ \bibinfo {author} {\bibfnamefont {Elcio}\ \bibnamefont
  {Abdalla}},\ }\bibfield  {title} {\enquote {\bibinfo {title} {{Stability of
  the curvature perturbation in dark sectors' mutual interacting models}},}\
  }\href {\doibase 10.1016/j.physletb.2008.11.062} {\bibfield  {journal}
  {\bibinfo  {journal} {Phys. Lett. B}\ }\textbf {\bibinfo {volume} {671}},\
  \bibinfo {pages} {139--145} (\bibinfo {year} {2009})},\ \Eprint
  {http://arxiv.org/abs/0807.3471} {arXiv:0807.3471 [gr-qc]} \BibitemShut
  {NoStop}%
\bibitem [{\citenamefont {Jackson}\ \emph {et~al.}(2009)\citenamefont
  {Jackson}, \citenamefont {Taylor},\ and\ \citenamefont {Berera}}]{2009}%
  \BibitemOpen
  \bibfield  {author} {\bibinfo {author} {\bibfnamefont {Brendan~M.}\
  \bibnamefont {Jackson}}, \bibinfo {author} {\bibfnamefont {Andy}\
  \bibnamefont {Taylor}}, \ and\ \bibinfo {author} {\bibfnamefont {Arjun}\
  \bibnamefont {Berera}},\ }\bibfield  {title} {\enquote {\bibinfo {title} {On
  the large-scale instability in interacting dark energy and dark matter
  fluids},}\ }\href {\doibase 10.1103/physrevd.79.043526} {\bibfield  {journal}
  {\bibinfo  {journal} {Physical Review D}\ }\textbf {\bibinfo {volume} {79}}
  (\bibinfo {year} {2009}),\ 10.1103/physrevd.79.043526}\BibitemShut {NoStop}%
\bibitem [{\citenamefont {Gavela}\ \emph {et~al.}(2010)\citenamefont {Gavela},
  \citenamefont {Lopez~Honorez}, \citenamefont {Mena},\ and\ \citenamefont
  {Rigolin}}]{Gavela:2010tm}%
  \BibitemOpen
  \bibfield  {author} {\bibinfo {author} {\bibfnamefont {M.~B.}\ \bibnamefont
  {Gavela}}, \bibinfo {author} {\bibfnamefont {L.}~\bibnamefont
  {Lopez~Honorez}}, \bibinfo {author} {\bibfnamefont {O.}~\bibnamefont {Mena}},
  \ and\ \bibinfo {author} {\bibfnamefont {S.}~\bibnamefont {Rigolin}},\
  }\bibfield  {title} {\enquote {\bibinfo {title} {{Dark Coupling and Gauge
  Invariance}},}\ }\href {\doibase 10.1088/1475-7516/2010/11/044} {\bibfield
  {journal} {\bibinfo  {journal} {JCAP}\ }\textbf {\bibinfo {volume} {11}},\
  \bibinfo {pages} {044} (\bibinfo {year} {2010})},\ \Eprint
  {http://arxiv.org/abs/1005.0295} {arXiv:1005.0295 [astro-ph.CO]} \BibitemShut
  {NoStop}%
\bibitem [{\citenamefont {Cen}(2001)}]{Cen:2000xv}%
  \BibitemOpen
  \bibfield  {author} {\bibinfo {author} {\bibfnamefont {Renyue}\ \bibnamefont
  {Cen}},\ }\bibfield  {title} {\enquote {\bibinfo {title} {{Decaying cold dark
  matter model and small-scale power}},}\ }\href {\doibase 10.1086/318861}
  {\bibfield  {journal} {\bibinfo  {journal} {Astrophys. J. Lett.}\ }\textbf
  {\bibinfo {volume} {546}},\ \bibinfo {pages} {L77--L80} (\bibinfo {year}
  {2001})},\ \Eprint {http://arxiv.org/abs/astro-ph/0005206}
  {arXiv:astro-ph/0005206} \BibitemShut {NoStop}%
\bibitem [{\citenamefont {Oguri}\ \emph {et~al.}(2003)\citenamefont {Oguri},
  \citenamefont {Takahashi}, \citenamefont {Ohno},\ and\ \citenamefont
  {Kotake}}]{Oguri:2003nn}%
  \BibitemOpen
  \bibfield  {author} {\bibinfo {author} {\bibfnamefont {Masamune}\
  \bibnamefont {Oguri}}, \bibinfo {author} {\bibfnamefont {Keitaro}\
  \bibnamefont {Takahashi}}, \bibinfo {author} {\bibfnamefont {Hiroshi}\
  \bibnamefont {Ohno}}, \ and\ \bibinfo {author} {\bibfnamefont {Kei}\
  \bibnamefont {Kotake}},\ }\bibfield  {title} {\enquote {\bibinfo {title}
  {{Decaying cold dark matter and the evolution of the cluster abundance}},}\
  }\href {\doibase 10.1086/378490} {\bibfield  {journal} {\bibinfo  {journal}
  {Astrophys. J.}\ }\textbf {\bibinfo {volume} {597}},\ \bibinfo {pages}
  {645--649} (\bibinfo {year} {2003})},\ \Eprint
  {http://arxiv.org/abs/astro-ph/0306020} {arXiv:astro-ph/0306020} \BibitemShut
  {NoStop}%
\bibitem [{\citenamefont {Malik}\ \emph {et~al.}(2003)\citenamefont {Malik},
  \citenamefont {Wands},\ and\ \citenamefont {Ungarelli}}]{Malik:2002jb}%
  \BibitemOpen
  \bibfield  {author} {\bibinfo {author} {\bibfnamefont {Karim~A.}\
  \bibnamefont {Malik}}, \bibinfo {author} {\bibfnamefont {David}\ \bibnamefont
  {Wands}}, \ and\ \bibinfo {author} {\bibfnamefont {Carlo}\ \bibnamefont
  {Ungarelli}},\ }\bibfield  {title} {\enquote {\bibinfo {title} {{Large scale
  curvature and entropy perturbations for multiple interacting fluids}},}\
  }\href {\doibase 10.1103/PhysRevD.67.063516} {\bibfield  {journal} {\bibinfo
  {journal} {Phys. Rev. D}\ }\textbf {\bibinfo {volume} {67}},\ \bibinfo
  {pages} {063516} (\bibinfo {year} {2003})},\ \Eprint
  {http://arxiv.org/abs/astro-ph/0211602} {arXiv:astro-ph/0211602} \BibitemShut
  {NoStop}%
\bibitem [{\citenamefont {Ziaeepour}(2004)}]{Ziaeepour:2003qs}%
  \BibitemOpen
  \bibfield  {author} {\bibinfo {author} {\bibfnamefont {Houri}\ \bibnamefont
  {Ziaeepour}},\ }\bibfield  {title} {\enquote {\bibinfo {title} {{Quintessence
  from the decay of a superheavy dark matter}},}\ }\href {\doibase
  10.1103/PhysRevD.69.063512} {\bibfield  {journal} {\bibinfo  {journal} {Phys.
  Rev. D}\ }\textbf {\bibinfo {volume} {69}},\ \bibinfo {pages} {063512}
  (\bibinfo {year} {2004})},\ \Eprint {http://arxiv.org/abs/astro-ph/0308515}
  {arXiv:astro-ph/0308515} \BibitemShut {NoStop}%
\end{thebibliography}%

\end{document}